\documentclass[11pt]{article}
\usepackage[nosort]{cite}
\usepackage{amsmath,amsthm,amsfonts,amssymb,amscd,mathrsfs,slashed,graphicx,braket}

\newlength{\xtrawidth}
\newlength{\xtraheight}
\numberwithin{equation}{section}
\numberwithin{table}{section}
\numberwithin{figure}{section}
\newcommand{\ii}{\mathrm{i}}
\newcommand{\dd}{\mathrm{d}}

\newcommand{\num}{{}}

\usepackage{xcolor}
\usepackage{arydshln}
\usepackage[font=footnotesize]{caption}
\usepackage{float}
\usepackage{braket}
\usepackage{extarrows}

\usepackage{bbm}
\usepackage{nicefrac}
 \usepackage{import}
 \usepackage{bbold}
\usepackage{mathrsfs}
\usepackage{pgf,tikz}
\usetikzlibrary{arrows}
\usepackage[fleqn]{mathtools}

\usepackage{graphicx}
\usepackage{subcaption}

\usepackage{dsfont}

\newcommand{\hoch}[1]{^{(#1)}}

\newenvironment{gleichung}{\begin{equation}\begin{aligned}}{\end{aligned}\end{equation}}
\newenvironment{gleichung*}{\begin{equation*}\begin{aligned}}{\end{aligned}\end{equation*}}

\renewcommand{\bf}{\textbf}

\newcommand{\QCO}{Quantum Differential Operator}

\begin{document}
\begin{titlepage}

\begin{center}
\hfill BONN--TH--2018--0X\\
\vskip 0.6in
{\LARGE\bf{WKB Method and Quantum Periods}\\\vspace{0.25cm} \bf{beyond Genus One}}

\vspace{20mm}
{
Fabian Fischbach\footnote{fischbach@physik.uni-bonn.de}, 
Albrecht Klemm\footnote{aklemm@th.physik.uni-bonn.de}, 
Christoph Nega\footnote{cnega@th.physik.uni-bonn.de }\\[4mm]
{\textit{\textsuperscript{123}Bethe Center for Theoretical Physics and \textsuperscript{2}Hausdorff Center for Mathematics,\\
{Universit\"{a}t Bonn, D-53115 Bonn}}} }\\[2mm]

\end{center} 

\vspace{2cm}
\begin{center}
\textbf{Abstract}
\end{center}
We extend topological string methods in order to perform WKB approximations for quantum 
mechanical problems with higher order potentials efficiently. This  requires techniques for the 
evaluation of the relevant quantum periods for Riemann surfaces beyond genus one. The 
basis  of these quantum periods is fixed using the leading behaviour of the classical periods. 
The full expansion of the quantum periods is obtained using a system of Picard-Fuchs like operators 
for a sequence of integrals of meromorphic forms of the second kind.  Discrete automorphisms 
of simple higher order potentials allow to view the corresponding higher genus curves as 
covering of a genus one curve. In this case the quantum periods can be alternatively obtained 
using the holomorphic anomaly solved in the holomorphic limit within the ring of quasi modular 
forms of a congruent subgroup of SL$(2,\mathbb{Z})$ as we check for a symmetric sextic potential.                   
 
\end{titlepage}

\section{Introduction and Summary}
Periods of differentials of first, second and third kind on Riemann surfaces $\Sigma_g$ 
with genus $g\ge 1$ are a classical mathematical subject generalizing the theory of elliptic functions.  
These periods  have rich applications to N=2 super symmetric gauge theory~\cite{Seiberg:1994rs,Klemm:1995wp},  
topological string theory on local Calabi-Yau manifolds~\cite{Katz:1996fh,Chiang:1999tz}, matrix models~\cite{Akemann:1996zr} 
and integrable models~\cite{Nekrasov:2009rc,Aganagic:2003qj}. Additional connections 
to Liouville Theory and more general 2d  CFTs have been  proposed in~\cite{Alday:2009aq} and 
at a more technical level the periods are  related to certain Feynman integrals~\cite{Bloch:2016izu}.   

The supersymmetric gauge theories exhibit a two parameter $\epsilon_1$, $\epsilon_2$ space of deformations,  the so called 
$\Omega$-background, that allows to solve it by localization~\cite{Nekrasov:2002qd}, 
while the topological string has generically only a world-sheet genus $g_\text{ws}$ expansion in the 
string coupling $g_s$ corresponding  to $g_s^2= \epsilon_1 \epsilon_2$. However, in the 
presence of global $U(1)_R$ symmetry on the geometry it can be uniquely refined motivically to exhibit two deformation 
parameters say $g_s$ and $s=(\epsilon_1+\epsilon_2)^2$~\cite{Choi:2012jz}. 
The matrix model has a genus expansion and also  a  candidate for refinement of the 
measure~\cite{Brini:2010fc}, while the particular integrable structure~\cite{Nekrasov:2009rc}
occurs in the Nekrasov-Shatashvili limit\footnote{We will denote  
the remaining deformation $\epsilon_1=\hbar$.} $\varepsilon_2=0$. The connections between these theories are 
mostly well understood. 
The N=2 super symmetric gauge theory is related to the topological string by the geometric 
engineering limit~\cite{Klemm:1995wp}. The relation between the topological string and 
the matrix models was made precise  in~\cite{Bouchard:2007ys}. 
 
It has been pointed out in~\cite{Aganagic:2003qj} that the genus expansion of the topological string can be viewed as a 
time dependent quantization of the geometry of the Riemann surface and in \cite{Aganagic:2011mi} 
it has been recognized that the  $\hbar$ expansion in the Nekrasov-Shatashvili limit (NS limit) can be literally 
viewed as a time independent WKB quantisation of an action given by the classical periods, which encode the
potential of the quantum mechanical problem.  This yields as solution to the WKB problem the $\hbar$ 
expansion of the  quantum periods.  More concretely it has been exemplified in~\cite{Codesido:2016dld} 
that solving the holomorphic anomaly equation can be turned into  an efficient formalism to obtain these quantum periods 
of simple WKB problems. This seems  independent of the fact whether $\Sigma_g$ corresponds topological string--,  
gauge theory-- or a matrix model spectral  curve. Quantum mechanical problems that do correspond to a gauge theory curve have been considered in \cite{Grassi:2018spf}. In particular  
near the Argyres-Douglas points for $SU(2)$ with one and two flavors in the fundamental  representation, 
the families of Seiberg-Witten curves describing the deformation away from the conformal point 
can be mapped exactly to the quantum mechanical problems with cubic and the quartic 
potential~\cite{Grassi:2018spf}.  As a consequence the quantized curve describes the gauge theory 
coupled to an $\Omega$-background.  A similar analysis could be done for the Argyres-Douglas 
points in the Coulomb branch of higher rank gauge groups.  

The holomorphic anomaly equations originate in the worldsheet B-model approach 
to topological string theory~\cite{Bershadsky:1993cx} which has only a $g_s$ expansion. 
For the present purpose they have to be refined as in~\cite{Huang:2010kf,Krefl:2010fm}. The
resulting equations are recursively based as starting data on the holomorphic genus zero free energy 
$F_{0}=F^{(0,0)}(t)$, which can be interpreted as the classical term, the an-holomorphic 
genus one free energy  $F^{(0,1)}(t)$, whose an-holomorphicity is given by a second  order  
$\partial_t \bar \partial_{\bar t}$ differential equation for the Ray-Singer torsion~\cite{Bershadsky:1993cx} giving 
$g_s^2$ corrections and a meromorphic function $F_{1}=F^{(1,0)}(t)$, which is proportional to the logarithm of the discriminant 
of $\Sigma_g$ and gives $\hbar^2$ corrections.  More generally the holomorphic anomaly equation for each 
$F^{(m,n)}(t)$ with $(m+n)\ge 2$ has a {\sl recursion kernel} called the holomorphic ambiguity, whose 
{\sl finite dimension} has a polynomial growth in $(n,m)$. It has been  argued in~\cite{Huang:2006si} 
that for gauge theory and matrix models this holomorphic ambiguity can be fixed by 
the gap condition at the conifold divisors and regularity at the orbifold divisors in the moduli space. 
In \cite{Haghighat:2008gw} it has been shown that this is more generally true for the genus expansion of 
the topological string on local geometries whose  mirrors are genus $g$ curves. These arguments 
have been extended to the  refined  theories whose  B-model description are genus $g$ curves. 
For this class the set of sufficient boundary conditions has been  specified  in \cite{Huang:2010kf,Huang:2011qx}.
They imply, of course,  the integrabiliy of  the theory in the NS limit, in which it is considerably simpler. 

In particular, one can obtain in these cases the quantum periods by solving the holomorphic
anomaly equation~\cite{Huang:2012kn}. Alternatively, one can consider the  quantum expansion of the 
 differential in $\hbar^2$, which is apart from the leading term in $\hbar$ always of the  
second kind and solve the quantum period using a system of  Picard-Fuchs like differential operators ${\cal D}_{2n}$ that 
yield the  $\hbar^{2n}$  correction term from the  classical period as proposed 
in~\cite{Huang:2014nwa}. 

In this paper  we extend both methods to higher genus curves. As it turns out the formalism using the holomorphic  
anomaly equation is slightly more convenient in the $g=1$ case, because one can use very efficiently 
the modularity properties of genus one curves  as developed in~\cite{Huang:2011qx}. One   
needs essentially only the genus one curve in the Weiererstrass form,  the transformation into which is easily done using Nagell's algorithm. This method  solves right away all cases with quantum mechanical problems
whose potential  is of quartic or cubic degree. However, we find that some  symmetric higher 
degree quantum mechanical potentials describe higher genus curves which are multi-coverings 
of genus one curves. In this case the higher genus problem can be reduced to a genus one problem  
and solved  as such very efficiently as described above. In particular, we check along these lines 
that the gap condition appropriately determines the right boundaries of the WKB problem with a sextic curve covering a quartic curve.  Higher degree potentials are of course  particularly interesting if this
symmetry can be broken at will by arbitrary  perturbations which  leads to really independent 
cycles or branch cuts which can capture interesting changes in the possible non-perturbative
effects. 

Despite the fact  that part of the original motivation  of this work was to use the extension of 
the SL$(2,\mathbb{Z})$ modular approach of the $g=1$ case to the one of Siegel modular forms 
in the higher genus case  in order to solve the holomorphic anomaly equation as in \cite{Klemm:2015iya}, 
it has turned out to be much easier to  extend the  ${\cal D}_{2n}$ to  a system of multi-parameter operators 
that yield the higher  $\hbar$ correction terms from the classical periods. We develop this formalism, which applies 
to arbitrary genus, and exemplify it with the quintic potential with additional deformation parameters turned on. 
Our ability to solve  quantum periods on higher genus curves  for arbitrary multi-parameter complex 
deformation families will also shed light  on the problem of how to restrict the general parametrization of higher genus curves 
by the Siegel upper  half plane to those  algebraic deformations of the potential that occur in a specific quantum mechanical setting, 
which we leave, however, for future explorations.%
Quantum periods could lead to a generalization of the
semiclassical analysis of 1D multivalent Coulomb gases mapped to  (non-)Hermitian quantum mechanics~\cite{Gulden:2013wya}, where the 
classical periods of families of Riemann surfaces of genus $g\ge 1$ yield information about the  energy spectrum and bandwidth (i.e. pressure and transport barrier).     

\noindent
{\bf {Note:}}  While we were preparing this preprint, the paper~\cite{Kreshchuk:2018qpf} 
appeared, which  has overlap in determining the quantum periods by the Picard-Fuchs differential systems, 
but only in the modular cases related to genus one curves, where the method was already discussed 
in~\cite{Huang:2014nwa} and the direct integration method is more efficient.


\section{The All-Orders WKB Method}

\subsection{The WKB Ansatz}

This section provides a quick introduction to the all-orders WKB method of Dunham \cite{PhysRev.41.713}\footnote{Our exposition closely follows \cite{Codesido:2016dld}. For a more pedagogical treatment we refer to \cite{galindo2}
.}. A central object in this method is the \textit{quantum period}, a formal power series in $\hbar^2$ whose $\hbar \rightarrow 0$ limit yields a phase space volume defined by a maximal energy $\xi$. Let us explain the construction. Consider the one-dimensional Schr\"{o}dinger equation for a particle moving in a potential $V$,
\begin{equation}\label{eq:schroedinger_eq}
{\hbar^2}\psi''(x) + p^2(x) \psi(x) = 0 ,\qquad p(x)=\sqrt{2(\xi-V(x))},
\end{equation}
where the mass has been set to $m=1$. The WKB ansatz for the wavefunction
\begin{equation}\label{eq:wkb_ansatz}
\psi(x) = \exp \left[  \frac{\ii}{\hbar} \int^x Q(x')\ \dd x' \right]
\end{equation}
turns the Schr\"{o}dinger equation into a Riccati equation for $Q(x)$,
\begin{equation}\label{eq:riccati_wkb}
Q^2(x) - \ii \hbar \frac{\dd Q(x)}{\dd x} = p^2(x),
\end{equation}
which in turn is expanded in a formal power series
\begin{equation}\label{eq:hbar_series}
Q(x) = \sum_{n=0}^\infty Q_n(x) \  \hbar^n
\end{equation}
and leads to a recursion for the functions $Q_n(x)$,
\begin{align}
Q_0(x) & = p(x) \nonumber \\
Q_{n+1}(x) & = \frac{1}{2Q_0(x)} \left( \mathrm{i} \frac{\partial}{\partial x}Q_n(x) - \sum_{k=1}^n Q_k(x)\ Q_{n+1-k}(x) \right).\label{eq:wkb_recursion} 
\end{align}
Splitting the series into even and odd powers of $\hbar$,
\begin{equation}
Q(x)= Q_\text{odd}(x) + P(x), \qquad P(x)=\sum_{n=0} Q_{2n}(x) \hbar^{2n},
\end{equation}
one finds that $Q_\text{odd}(x)$ is a total derivative
\begin{equation}
Q_\text{odd}(x) = \frac{\ii \hbar}{2} \frac{P'(x)}{P(x)} =  \frac{\ii \hbar}{2} \log P(x)
\end{equation}
and it follows that only $Q_1$ and $P$ contribute to period integrals of \eqref{eq:hbar_series}. The WKB wavefunction \eqref{eq:wkb_ansatz} can be rewritten in terms of $P(x)$ only, i.e.,
\begin{equation}
\psi(x) = \frac{1}{\sqrt{P(x)}} \exp\left[ \frac{\ii}{\hbar} \int^x P(x') \ \dd x' \right].
\end{equation}

Now consider the family of (compact) Riemann surfaces $\Sigma$ defined by
\begin{equation}\label{eq:wkb_curve_general}
\Sigma : \quad y^2 = p^2(x),
\end{equation}
where the energy $\xi$ and possibly parameters of the potential $V$ serve as family parameters. In the following we restrict ourselves to polynomial potentials such that $\Sigma$ becomes a family of hyperelliptic curves. The genus $g=\lfloor \frac{d-1}{2}\rfloor$ is then determined by the degree $d$ of the potential.

 For anharmonic oscillators leading to a genus one curve there are only two independent one-cycles on the curve, one of which typically encircles the branch points bounding a classically allowed region. We call it the $A$-cycle. On the other hand, the $B$-cycle corresponds to a classically forbidden region. The volume of the classically allowed region in phase space can be expressed as the period integral of the one-form $y(x) \ \dd x$ as 
\begin{equation}
\frac{1}{\pi} \operatorname {vol}_0(\xi) = \frac{1}{2\pi}\oint_A y(x) \ \dd x.
\end{equation}
Its quantum counterpart is then defined using the formal series for $P(x)\ \dd x$,
\begin{equation}
\nu = \frac{1}{2\pi}\oint_A P(x) \ \dd x,
\end{equation}
and appears in the all-orders WKB quantization condition
\begin{equation}
\nu = \hbar \left(n + \frac{1}{2} \right), \qquad n=0,1,2, \dots \ .
\end{equation}
The questions of how to generalize this quantization condition to (1.) several potential wells separated by barriers (as it is generic for higher genus curves) and to (2.) include non-perturbative effects shall not be adressed here. Indeed, we will only be interested in providing efficient computational methods for WKB quantum periods and in  probing the conjectural connection between these periods and the holomorphic anomaly equation.

\subsection{Geometry of Quantum Periods}
\label{sec_geometry}

In the following we collect some facts\footnote{The mathematical statements we have collected here are mostly textbook knowledge. A good introduction to Riemann surfaces with application to periods is given in \cite{donaldson2011riemann}.} about the geometry and topology of hyperelliptic curves, tailored to the WKB setup and yielding explicit constructions.

\paragraph{Moduli Space of Hyperelliptic Curves.}
A compact Riemann surface of genus $g\geq2$ has $\dim_\mathbb{C} \mathcal{M}_g = 3g-3$ moduli, while the moduli space of hyperelliptic curves is of dimension $\dim_\mathbb{C} \mathcal{M}^\mathrm{hyp}_g=2g-1$. 
%
%
To understand the latter, note that the branch points serve as local parameters of $\mathcal{M}^\mathrm{hyp}_g$. Two hyperelliptic (or elliptic) curves are conformally equivalent if and only if their branch points differ by a fractional linear transformation  (M\"{o}bius transformation)
\begin{equation}
x \longmapsto \frac{\alpha x + \beta}{\gamma  x + \delta}~, \qquad 
\begin{pmatrix}
\alpha & \beta \\
\gamma & \delta
\end{pmatrix} \in \mathrm{PSL}(2,\mathbb{C}).
\end{equation}
These are precisely the bijective holomorphic maps $\hat{\mathbb{C}} \rightarrow \hat{\mathbb{C}}$. Hence $\dim_\mathbb{C} \mathcal{M}^\mathrm{hyp}_g=(2g+2)-\dim_\mathbb{C} \mathrm{PSL}(2,\mathbb{C})=2g-1$ as claimed.

From the WKB point of view the energy $\xi$ is a natural curve parameter and a choice of the potential $V$ selects a sublice of $\mathcal{M}^\mathrm{hyp}_g$.

\begin{figure}
  \centering
  \includegraphics[width=\textwidth]{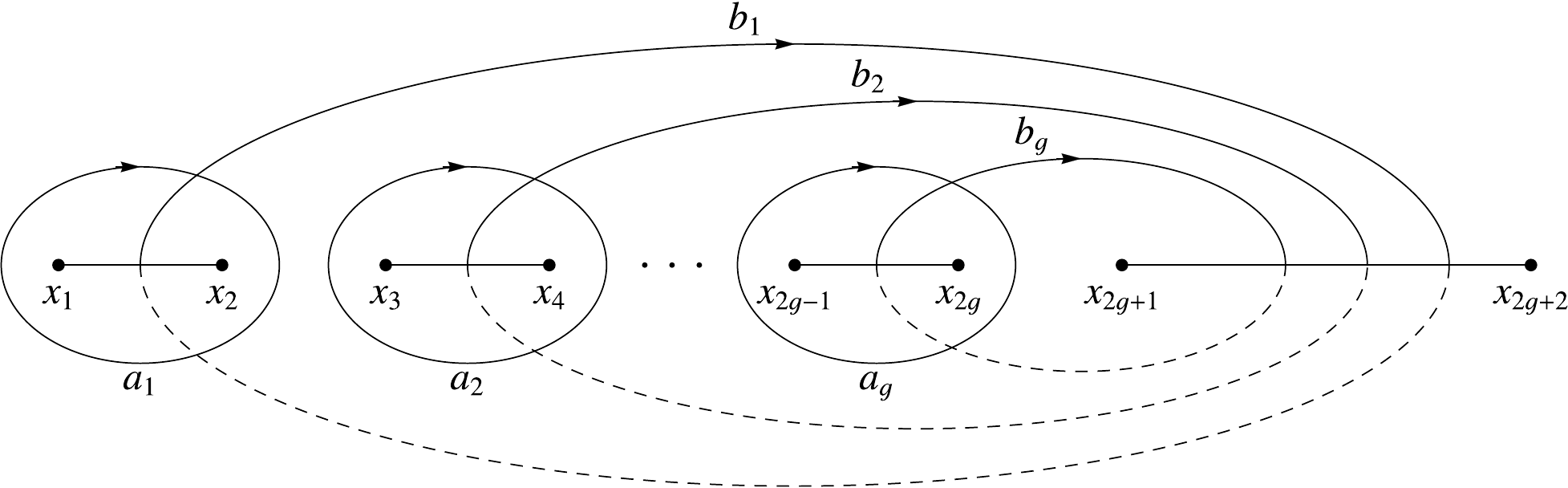}
  \caption[A canonical basis of one-cycles for a hyperelliptic curve.]{A canonical basis of one-cycles for a hyperelliptic curve. Branch points $x_j$ are algebraic functions of $\xi$. Dotted path segments lie on the other sheet.}
  \label{fig:branch_cuts}
\end{figure}

\begin{figure}[htbp]
  \centering
  \includegraphics[width=0.75\textwidth]{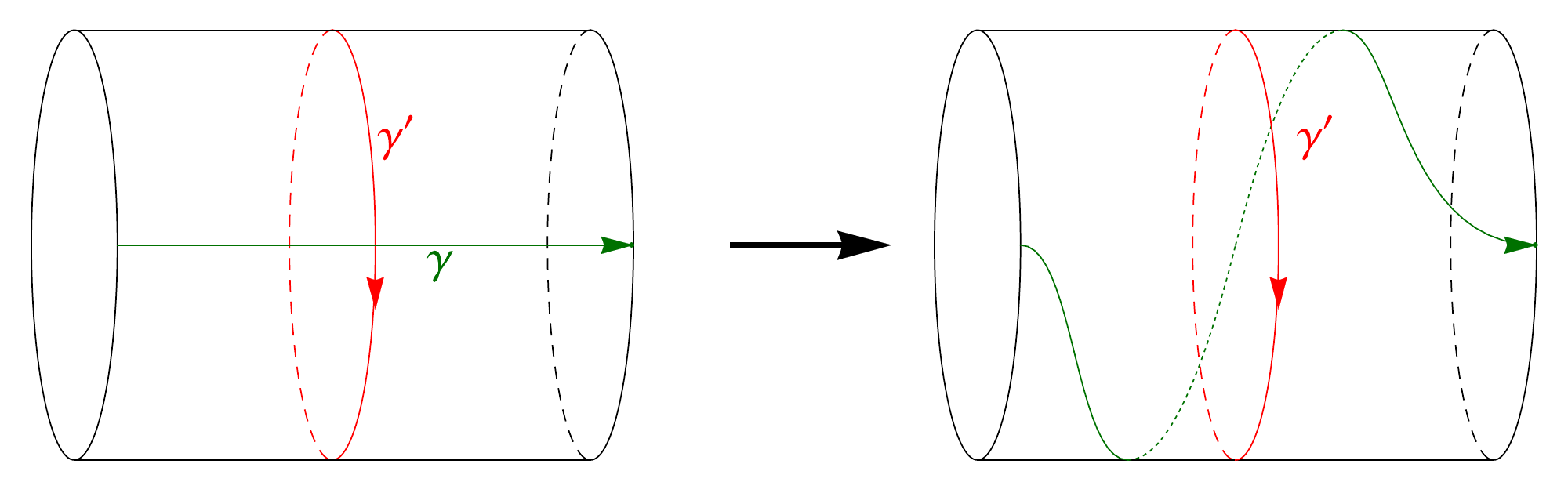}
  \caption[Dehn twist along an embedded circle.]{Dehn twist along an embedded circle $\gamma'$.}
  \label{fig:dehn_twist}
\end{figure}

\paragraph{Degenerations and Monodromies.}
For a polynomial $V(x)$ the complex curve $\Sigma$ can be regarded as two copies of the Riemann sphere $\hat{\mathbb{C}}=\mathbb{P}^1$,  glued together along branch cuts. Here the branch points are turning points of classical trajectories and their complex analogs\footnote{If $V(x)$ has odd degree, one branch point lies at infinity.}, given by $p^2(x)=0$. There are $\dim_\mathbb{C} H_1(\Sigma,\mathbb{C})=2g$ independent one-cycles on $\Sigma$, which can be represented by closed contours encircling branch points. This is schematically shown in Fig. \ref{fig:branch_cuts}.
For certain values of $\xi$ two (or more) branch points coincide and a corresponding one-cycle vanishes. Pictorially, a handle of the surface described by $\Sigma$ gets pinched and $\Sigma$ is called degenerate or singular. Indeed, the curve is non-singular if and only if the discriminant
\begin{equation}
\operatorname{discr}_x p^2(x)= c \prod_{i<j}(x_i - x_j)^2
\end{equation}
is non-zero. This is always a polynomial in the coefficients multiplying the monomials of $p^2(x)$. Generically, there are no radical expressions for the roots $x_i$ if $d>5$ (Abel-Ruffini), nevertheless the discriminant can be factorized as\footnote{This can be seen as follows: up to a constant the discriminant is the resultant of $(\xi-V(x))$ and $V'(x)$, which vanishes if and only if the two posses a common root.}
\begin{equation}\label{eq:disc_factorization}
\Delta(\xi) = \prod_{i=1}^{d-1} \left(\xi - V(x^{(c)}_i) \right)
\end{equation}
where the $x^{(c)}_i$ are the critical points of $V(x)$. Thus, for fixed potential $V(x)$, eq. \eqref{eq:wkb_curve_general} defines a smooth fiber bundle $\mathcal{S} \stackrel{\pi}{\rightarrow} \mathcal{B}$ over the base $\mathcal{B} = \mathbb{P}^1 \backslash (\lbrace \Delta = 0 \rbrace \cup \lbrace \infty \rbrace)$, the fibers being smooth hyperelliptic curves. Moreover, the homology groups $H_1(\Sigma,\mathbb{C})$ of the fibers combine to a complex vector bundle over the same base $\mathcal{B}$. Fixing a reference point $\xi$ in $\mathcal{B}$ for the moment, it can be shown \cite{donaldson2011riemann} that there is a monodromy homomorphism\footnote{The fundamental group of the base appearing here is generated by loops starting at $\xi$ and encircling single points of $\lbrace \Delta = 0 \rbrace \cup \lbrace \infty \rbrace$ in some fixed direction once (see Fig. \ref{fig:moduli_sphere_quintic} for an example).} 
\begin{equation}\label{eq:monodromy_homom}
\rho_\mathrm{mon.} : \   \pi_1\left(\mathcal{B},\ \xi \right)   \rightarrow  \mathrm{GL}\left(H_1\left(\Sigma_\xi, \mathbb{C}\right)\right)
\end{equation}
whose image can be identified with a discrete subgroup of $\mathrm{Sp}\left(2g, \mathbb{Z}\right)$, i.e., it preserves the symplectic structure defined by the intersection form $\cap$. The image of \eqref{eq:monodromy_homom} is the \textit{monodromy group} of the family of curves. %
  For the generator $g'$ associated to a point $\xi'$ with vanishing cycle $\gamma'$ the monodromy action on a cycle $\gamma \in H_1\left(\Sigma_\xi, \mathbb{C}\right)$ is given by the Picard-Lefshetz formula
\begin{equation}\label{eq:picard-lefshetz}
\rho_\mathrm{mon.}(g') \ : \quad   \gamma \mapsto \gamma + \left(\gamma'\cap\gamma\right)\ \gamma'.
\end{equation}
Geometrically this is a Dehn twist along the embedded circle representing $\gamma'$. There is a neighborhood $N$ of this circle homeomorphic to a cylinder $[-1,1]\times \mathbb{S}^1$. If $\gamma$ is another embedded circle intersecting $\gamma'$ exactly once with the intersection point lying in $N$, then Fig. \ref{fig:dehn_twist} shows the action corresponding to \eqref{eq:picard-lefshetz}.
The upshot is that this monodromy determines the structure of the periods $\Pi(\xi)$ on $\Sigma$, which are sections of the same vector bundle over $\mathcal{B}$ (up to isomorphism). This will become explicit when discussing Picard-Fuchs equations shortly.

\paragraph{Abelian Differentials.} The WKB periods belong to a tower of meromorphic one-forms $\mathcal{Q}_n = Q_n(x) \ \dd x$ on $\Sigma$. Meromorphic one-forms on Riemann surfaces are usually divided into three types: \textit{Abelian differentials of the first kind} are holomorphic, i.e., in any local complex coordinate $(z,U)$ they can be written as $\lambda=\lambda_z \ \dd z$ with a holomorphic function $\lambda_z$. For hyperelliptic curves of genus $g$ there are precisely $g$ such forms modulo exact forms, a basis is given by the $g$ differentials  $x^j \dd x /y \ $, $j=0,...,g-1$. There are also  \textit{Abelian differentials of the second kind} where $\lambda_z$ is a meromorphic function with vanishing residues. They form an infinite-dimensional vector space $\Lambda^1(\Sigma)$. An \textit{Abelian differential of the third kind} is also allowed to have non-zero residues in its local Laurent expansion.

It is well known that the middle cohomology $H^1(\Sigma, \mathbb{C})$ has complex dimension $2g$ and can be represented as the quotient
\begin{equation}\label{eq:H1_quotient}
H^1(\Sigma, \mathbb{C}) = \frac{\Lambda^1(\Sigma)}{\text{Im} \left(  \operatorname{d} : \Lambda^0(\Sigma) \rightarrow \Lambda^1(\Sigma) \right)},
\end{equation}
where $\operatorname{d}$ denotes the exterior derivative and we have introduced $\Lambda^0(\Sigma) = \lbrace \phi : \Sigma \rightarrow \mathbb{P}^1 \ \vert \ \phi \ \text{meromorphic} \rbrace $. Clearly  non-zero residues obstruct homotopy invariance, which is required for a well-defined pairing between singular homology $H_1(\Sigma,\mathbb{C})$ and cohomology $H^1(\Sigma,\mathbb{C})$. Some computation will be required to see which WKB differentials are of third or second kind.



\paragraph{Picard-Fuchs Equations.} Periods of differentials of the second kind satisfy ordinary differential equations with respect to the modulus $\xi$, which are at most of order $2g$. To see this, recall that any period and its derivatives produce local sections of a vector bundle over $\mathcal{B}$, the fibers being isomorphic to $H^1(\Sigma, \mathbb{C})$. Finite dimensionality then requires a linear relation amongst the derivatives. These \textit{Picard-Fuchs equations} (PFE) are of Fuchsian type and fundamental systems can be constructed by Frobenius' method \cite{ince1956ordinary}.%

To find the PFE for the WKB periods at each order in $\hbar^2$ we use the identification \eqref{eq:H1_quotient} and start by considering the WKB differentials. For each $n \in\lbrace 0,1,2,...\rbrace$ there is a polynomial $p_n$ in $x$ and $\xi$ of degree
\begin{equation}\label{eq:deg_x_pn}
 \deg_x p_n \leq n(d-1) 
\end{equation}
in $x$ 
 such that
\begin{equation}\label{eq:wkb_Qn_polyn}
Q_n = \frac{p_n}{y^{3n-1}}.
\end{equation}
%
 Polynomials for $n\geq 2$ satisfy the recursion relation
\begin{equation}\label{eq:wkb_pn_rec}
p_{n+1} = \frac{\ii}{2}\left(2  p_n' \ (\xi - V) + (3n-1) p_n V' \right)- \frac{1}{2} \sum_{k=1}^n p_k p_{n+1-k}, \qquad p_0 = 1,\quad p_1=\frac{-\ii}{2}V'
\end{equation}
where $(...)'$ means $\partial_x(...)$. As $V$ is a real polynomial, $p_{2n}$ ($p_{2n+1}$) has real (imaginary) coefficients. Linear dependence in $H^1(\Sigma, \mathbb{C})$ hence translates into the ansatz
\begin{equation}\label{eq:pfe_cohom_ansatz}
\sum_{i=0}^{r} f_i(\xi) \ \frac{\partial^{i} }{ \partial\xi^{i}} \left( \frac{p_n}{y^{3n-1}}\right)- \sum_{k=0}^{k_\mathrm{max}}  \  \frac{\partial}{\partial x}\left(\alpha_k(\xi) \frac{x^{k}}{y^{3n-3+2r}}\right) = 0
\end{equation}
for some $r\leq 2g$ and sufficiently large $k_\text{max}$. After writing the expression with $y^{3n-1+2r}$ as common denominator, we choose the $\alpha_k$ to subsequently eliminate monomials $g_j(\xi) x^j$ from the resulting numerator, starting from the highest power in $x$. Some monomials will remain and requiring their coefficients to vanish determines the polynomials $f_i(\xi)\in \mathbb{Z}[\xi]$ up to an overall constant. Thus, we find the PFE
\begin{equation}\label{eq:PFE}
\sum_{i=0}^{r} f_i(\xi) \ \frac{\partial^{i} \Pi(\xi) }{ \partial\xi^{i}}   = 0 \qquad \text{for} \qquad \Pi(\xi) = \oint_\gamma Q_n \ \dd x.
\end{equation}

The polynomial in front of the highest derivative is actually the discriminant $\Delta$ encountered before, i.e., degenerations of $\Sigma$ are in one-to-one-correspondence with the regular singular points of the PFE. The relation to Picard-Lefshetz theory becomes even clearer when regarding the analytic structure of solutions to the PFE \eqref{eq:PFE}: the series ansatz
\begin{equation}\label{eq:pfe_ansatz}
\Pi(\xi) = \xi^r \sum_{k=0}^\infty c_k \xi^k
\end{equation}
leads to a polynomial \textit{indicial equation} for $r$\footnote{The exponent $r$ appearing in this ansatz is not to be confused with the order $r$ of the Picard-Fuchs equation \eqref{eq:PFE}.}. Here we seek solutions around $\xi=0$. After having solved for $r$ a recursion relation for the coefficients $c^{(r)}_k$ of each $r$-solution is to be found. If the roots of the indicial polynomial are pairwise distinct and no two of them differ by an integer, this construction already leads to a fundamental system. In case two roots differ by an integer, without loss of generality $r_1-r_2\in \lbrace 0, 1, 2, ...\rbrace$ and $\Pi_{r_1}$ is a solution of the form \eqref{eq:pfe_ansatz}, a linearly independent solution may be found using the ansatz
\begin{equation}\label{eq:pfe_ansatz_log}
\Pi_{r_2}(\xi) = c \cdot \Pi_{r_1}(\xi)\ \log(\xi) + \sum_{k=0}^\infty d_k \ \xi^{k+r_2}
\end{equation}
where the constant $c$ might be zero if $r_1-r_2 \neq 0$\footnote{For a more complete discussion of Frobenius' method the reader may wish to  consult \cite{bender1978advanced} or \cite{ince1956ordinary}.}.
Consider the special case of a double indicial root $r_1 = r_2 = 1$, so the analytic solution $\Pi_1(\xi)=\xi + ...$ vanishes as $\xi \rightarrow 0$. The second solution $\Pi_{1'}$ for the given index $r=1$ takes the form \eqref{eq:pfe_ansatz_log} with $c\neq 0$, without loss of generality $c=(2 \pi \mathrm{i})^{-1}$, and thus possesses the monodromy property encountered in the Picard-Lefshetz formula: analytic continuation along a closed path enclosing $\xi=0$ shifts $\Pi_{1'} \rightarrow \Pi_{1'} + \Pi_{1} $ due to the logarithm. This monodromy property allows us to identify $\Pi_{1}$ with the period integral of the respective one-form along a one-cycle that vanishes as $\xi \rightarrow 0$ (up to normalization). Also, $\Pi_{1'}$ belongs to periods along cycles non-trivially intersecting with the vanishing cycle.

 
 \paragraph{Differential Operators for Quantum Corrections.} Exploiting linear dependence modulo exact forms we also obtain differential operators $\hbar^{2n} \mathcal{D}_{2n}$ that generate quantum corrections when acting on classical WKB periods $\oint Q_0 \ \dd x$. Derivatives of the latter span a subspace of $H^1(\Sigma, \mathbb{C})$, in all examples considered the full space, so an ansatz analogous to eq. \eqref{eq:pfe_cohom_ansatz} is justified. In the above fashion we find rational functions $q_i^{(2n)}(\xi)\in \mathbb{Q}(\xi)$ such that
 \begin{equation}\label{eq:quantum_diff_op}
 \mathcal{D}_{2n}Q_{0} := \left[ \sum_{i=0}^{2g-1} q_i^{(2n)}(\xi) \ \frac{\partial^i}{\partial \xi^i}\right] Q_0 = Q_{2n} + \partial_x (...).
 \end{equation}

We stress that these results provide an efficient method of computing WKB periods up to arbitrary order in $\hbar^2$. First one computes the classical PFE and uses Frobenius' method to provide a fundamental system of solutions. The difficult point is to compute a few terms\footnote{In appendix \ref{sec_evperiodint} we show how to compute the leading behaviour of the classical periods by means of an example.} of the classical hyperelliptic periods $\oint y \ \dd x$ in order to fix the corresponding linear combination of solutions. Applying $\hbar^{2n} \mathcal{D}_{2n}$ on those bypasses further evaluation of period integrals.


\section{Holomorphic Anomaly Equation and Direct Integration}
\label{sec_holanomalyeq}

\subsection{Connecting WKB to the Holomorphic Anomaly Equation}
\label{subsec:connection}

Having introduced the WKB method and discussed the geometry behind its periods we are now able to connect the latter to the holomorphic anomaly equation governing refined topological string free energies. Guided by \cite{Codesido:2016dld} we take from the WKB recursion \eqref{eq:wkb_recursion} the formal power series
\begin{gleichung}
	P(x) 	=	\sum_{n=0}^\infty Q_{2n}(x)\hbar^{2n}~.
\label{formal}
\end{gleichung}\noindent
In the examples discussed in \cite{Codesido:2016dld} there is a canonical symplectic pair $(A,B)$ of cycles, called the perturbative and non-perturbative cycle, encircling pairs of branch points localized on the real axis. With these cycles we define the quantum $A$- and $B$-period
\begin{align}
		\nu(\xi)	&=	\sum_{n=0}^\infty \nu^{(2n)}\hbar^{2n} =	\frac{1}{2\pi}\sum_{n=0}^\infty \hbar^{2n} \oint_A	 Q_{2n}~\mathrm dx \label{defaperiod}\\
	\nu_D(\xi)	&=	\sum_{n=0}^\infty \nu_D^{(2n)} \hbar^{2n}	 =	-i\sum_{n=0}^\infty \hbar^{2n}	\oint_B Q_{2n}(x)~\mathrm dx~.
\label{defbperiod}
\end{align}
We call the inverse of the $A$-period \eqref{defaperiod} the quantum mirror map\footnote{The leading term, the \textit{classical} mirror map $\xi_0(\nu)$, is the inverse function of the classical $A$-period.}
\begin{gleichung}
	\xi(\nu)	=	\sum_{n=0}^\infty \xi_n(\nu)\hbar^{2n}~.
\label{quantummirrormap}
\end{gleichung}\noindent
Then the quantum free energy is defined by
\begin{gleichung}
	\frac{\partial F}{\partial\nu}	=	\nu_D
\label{deffreeenergies}
\end{gleichung}\noindent
which makes it obvious that the quantum free energy admits an $\hbar$ expansion
\begin{gleichung}
	F(\nu)	=	\sum_{n=0}^\infty F_n(\nu)\hbar^{2n}~.
\end{gleichung}\noindent
As $F(\nu)$ was defined via its $\nu$-derivative it is not possible to fix the constant term of the free energy from the WKB perspective\footnote{The authors of \cite{Codesido:2016dld} however discuss a preferred choice of the constant term, which allows them to write down an integrated form of what is called a PNP-relation for the quantum mirror map specific to the respective systems investigated there.}.

Inspired by this construction Codesido and Marino \cite{Codesido:2016dld} claimed that free energies \eqref{deffreeenergies} corresponding 
to all-orders WKB periods \eqref{defaperiod} and \eqref{defbperiod} of generic one-dimensional quantum systems are governed by the 
refined holomorphic anomaly equation characterizing the refined topological string free energies in the Nekrasov-Shatash-vili limit.

\subsection{Sub-slices in the Moduli Space of Hyperelliptic Curves }
\label{subsec:subslices}

On  a genus $g$ Riemann surface $\Sigma_g$ it is always possible to choose a symplectic basis 
$(A_i,B^i)$, $i=1,\ldots,g$, of the homology $H_1(\Sigma_g,\mathbb{Z})$ with $A_i \cap B^j=\delta_i^j$ 
and accordingly one has  in general several $A$-periods $\nu_i(\xi, {\underline \eta})$ and several $B$-periods $\nu_{D\, i}(\xi, {\underline \eta})$ respectively. Moreover,  on a hyperelliptic curve $\Sigma_g$  they can in the general case depend, besides of the energy 
$\xi$ corresponding to the constant term in $x$, on $2g-2$ perturbations $\underline \eta $ of 
the potential.

Easy quantum mechanical  potentials  correspond to sub-slices in 
the full moduli space where the branch points of one cut are real.  
In the simplest cases the periods on this sub-slice enjoy a simple 
relation due to additional symmetries on $\Sigma_g$. For example for a sextic 
potential that we consider in Section \ref{subsec_sextic} two  symmetric 
$B$ cycles  are exchanged by $\mathbb{Z}_2$ symmetry and the problem 
reduces to one pair of symplectic periods on a genus one curve 
on which a solution of the holomorphic anomaly, described in some generality in 
Section \ref{subsection_rhae}-\ref{subsection_Fha}, can be constructed and perfectly 
reproduces the quantum period, just using the standard boundary conditions 
to fix the recursion kernels, as summarised for the particular example in 
Section \ref{subsec_onedimholeq}. 
More conceptually one can state that is this case if one can find a subgroup 
$\Gamma_{\Sigma_1}\subset {\rm SL}(2,\mathbb{Z})$ with $ \Gamma_{\Sigma_1}
\subset  \Gamma_{\Sigma_g} \subset {\rm Sp}(4 ,\mathbb{Z})$ on the sub-slice and 
the generators of almost modular forms of  $\Gamma_{\Sigma_1}$ can be 
used to perform the direct integration.    

The situation is more complicated if one considers a sub-slice on a 
higher genus surface when the reality condition of the branch points of one 
cut is not related to a symmetry, as for example for the particular 
quintic potential discussed in section \ref{subsec_quintic}. In this case it is not 
possible to obtain the solution of the holomorphic anomaly equation by 
the direct integration directly on the sub-slice. The 
reason is that the reality condition  breaks in general the  modular  invariance 
of the $F^{(n,g)}$ under the subgroup $\Gamma_{\Sigma_g} \subset {\rm Sp}(2 g ,\mathbb{Z})$ 
of the family $\Sigma_g$ in a very complicated way~\footnote{One can speculate 
that there is a subgroup of ${\rm SL}(2,\mathbb{R})$ which plays the role of 
the modular group on the sub-slice, but this would require to understand  
the relation of the fourth order differential equation (\ref{eq:pfe_classical_quintic}) 
to a Schwarzian system  along the lines discussed in~\cite{Candelas:1990rm}.}.  A modular family is 
however necessary to  write the  $F^{(n,g)}$ as polynomials in the 
ring of modular generators for $\Gamma_{\Sigma_g}$ and to perform the 
direct integration with respect to the almost holomorphic generators. Also in order to 
determine all boundary data by  the gap condition one has to consider in 
general all possible conifold divisors and some of those might not be 
accessible in the restricted parametrization of the sub-slice.  
At the technical level we argue in Section \ref{subsubsec_quinticF1} that 
on the sub-slice of the quintic one does not find the start datum $F_1$ 
that allows to setup the direct integration on the slice directly.       

These problems have a conceptually easy, but technically demanding 
solution: one has  to set up  the problem first in the full moduli space of the 
hyperelliptic curve, as it has been done e.g. for $g=2$ in~\cite{Klemm:2015iya}, 
and then restrict to the sub-slice.   

Before setting up the formalism to solve the multi-parameter holomorphic
anomaly equation we should mention that we develop a formalism to get 
the quantum period on the sub-slice  and the full moduli space by deriving 
a system of differential operators ${\cal D}_{2n}$  that allow to obtain all 
quantum periods if the classical periods over a symplectic basis are determined.
We demonstrate this for  the quintic in section~\ref{subsec_quintic}.


\subsection{The Refined Holomorphic Anomaly Equation}
\label{subsection_rhae} 
The refined holomorphic anomaly equation for the topological string~\cite{Huang:2010kf,Klemm:2015iya} is given by
\begin{gleichung}
	\bar\partial_{\bar \imath} F^{(n_1,n_2)} = \frac{1}{2} \bar C_{\bar \imath}^{jk} \left( D_j D_k F^{(n_1,n_2-1)} + \sum_{m,h}{\vphantom{\sum}}'  D_j F^{(m,h)}\cdot D_k F^{(n_1-m,n_2-h)}	\right)~
\label{refinedanomaly}
\end{gleichung}\noindent
for $n_1 + n_2>1$, where the prime mark indicates omission of terms with $(m,h)=(0,0)$ and $(n_1,n_2)$. The indices $i,j,k$ run over the number of different moduli  of the curve. Covariant derivatives $D_i$ correspond to the metric $G_{ij}$ on the moduli space of complex structures of $\Sigma_g$. The metric can be expressed 
using the standard period matrix $\tau_{ij}$ of the standard holomorphic one-forms 
of  $\Sigma_g$ as
\begin{gleichung}
	G_{i\bar{j}}	=	-i(\tau - \bar\tau)_{ij}~.
\label{metric}
\end{gleichung}\noindent
Furthermore, the expression $\bar C_{\bar \imath}^{jk}$ is related to the Yukawa coupling
\begin{gleichung}
	C_{ijk}	=	\frac{\partial^3 F^{(0,0)}}{\partial t_i\partial t_j\partial t_k}
\label{yukawa}
\end{gleichung}\noindent
by
\begin{gleichung}
	\bar C_{\bar \imath}^{jk}	=	G^{l\bar p}G^{m\bar n}\bar C_{\bar p \bar n \bar k}~.
\label{nonholyukawa}
\end{gleichung}\noindent
Moreover, the Yukawa coupling can be given for example by
\begin{gleichung} 
\frac{\partial \tau_{ij}}{\partial \nu_k}=C^m_i C^n_j C_{knm},
\end{gleichung}   
where the invertible constant matrix $C^m_i$ is given by a relative normalisation of the 
classical periods of the holomorphic one form differentials relative to the periods 
$\nu_i(\xi, {\underline \eta})$ and $\nu_{D\, i}(\xi, {\underline \eta})$ as discussed 
in~\cite{Klemm:2015iya}. This means that the $C_{\bar \imath}^{jk}$ are given entirely in 
terms of the classical periods.       

In \eqref{refinedanomaly} the anomaly equation is stated in its most general form. For the conjecture we have to restrict to the Nekrasov-Shatashvili (NS) limit~\cite{Huang:2012kn} $n_2 =0$ for which the first term of \eqref{refinedanomaly} drops out and we are left with the free energies
\begin{gleichung}
	F_n = F^{(n,0)}.
\label{nsfreeenergies}
\end{gleichung}\noindent


\subsection{Direct Integration Procedure in Terms of Propagators}
\label{subsection_DI}
Solving the anomaly equation can be done with the direct integration procedure\cite{Huang:2010kf,Huang:2006si,Klemm:2015iya}. Here we will use the direct integration method from ~\cite{Haghighat:2008gw}, which is formulated in terms of the propagator $S^{ij}$, the latter being defined by
\begin{gleichung}
	\bar\partial_{\bar k} S^{ij} = \bar C^{ij}_{\bar k}~.
\label{defpropagator}
\end{gleichung}\noindent  
The idea of the direct integration method is to rewrite the anti-holomorphic derivatives as derivatives with respect to the propagator
\begin{gleichung}
	\bar\partial_{\bar i} F^{(n_1,n_2)} = \bar C_{\bar i}^{jk} \frac{\partial F^{(n_1,n_2)}}{\partial S^{jk}}
\label{ideadirect}
\end{gleichung}\noindent
such that \eqref{refinedanomaly} becomes\footnote{Here we have to assume that the $\bar C_{\bar i}^{jk}$ are linearly independent. Furthermore, the covariant derivatives become normal derivatives because in the local case the K\"ahler connection in the covariant derivatives becomes trivial.}
\begin{gleichung}
	\frac{\partial F^{(n_1,n_2)}}{\partial S^{jk}} = \frac{1}{2}\left( D_j\partial_k F^{(n_1,n_2-1)} + \sum_{m,h}{\vphantom{\sum}}' \partial_j F^{(m,h)}\cdot \partial_k F^{(n_1-m,n_2-h)} \right)~.
\label{simplyfiedanomaly}
\end{gleichung}\noindent
The propagator can be calculated from a set of equations derived from special geometry of the moduli space~\cite{Klemm:2015iya}
\begin{gleichung}
	D_i S^{kl}		&=	-C_{imn} S^{km}S^{lm} + f_i^{kl}~,\\
	\Gamma_{ij}^k	&= - C_{ijl}S^{kl} + \tilde f_{ij}^k~,\\
	\partial_i F^{(0,1)}	&=	\frac{1}{2} C_{ijk}S^{jk} A_i~.
\label{equationsprop}
\end{gleichung}\noindent
These equations are overdetermined in the multi-moduli case. In the one-parameter case we solve these equations by imposing $A_\xi = 0$. The other two ambiguities can then be fixed. As an ansatz for the unknowns in \eqref{equationsprop} one can choose
\begin{gleichung}
	f_i^{kl}	=	\frac{h(\xi)}{\prod_r\Delta_r^p}~,
\label{ansatzambiguities}
\end{gleichung}\noindent
where $\Delta_r$ are different factors of the discriminant and $h(\xi)$ is some polynomial in the modulus $\xi$. An analog ansatz is made for $\tilde f_{ij}^k $.

As initial values for the holomorphic anomaly equation a suitable form for $F^{(0,0)}$, $F^{(1,0)}$ and $F^{(0,1)}$ is demanded. From special geometry $F^{(0,0)}$ can be determined. For $F^{(1,0)}$ and $F^{(0,1)}$, respectively, we take~\cite{Klemm:2015iya}
\begin{gleichung}
	F^{(1,0)}	&=	\frac{k_1}{24} \log\left(	\Delta~\xi^\alpha	\right)\qquad\text{and} \\
	\mathcal F^{(0,1)}	&=	\frac{k_1^\prime}{2} \log\left(	\Delta^a~\xi^b\cdot \left(	\frac{\partial t}{\partial \xi}	\right)^{-1}	\right)
\label{ansatzfreeenergies}
\end{gleichung}\noindent
as a suitable ansatz where the parameters $\alpha$, $a$ and $b$ have to be fixed. By comparison with the WKB results we can fix these parameters. 
With $\mathcal F^{(0,1)}$ we refer to the holomorphic part of $F^{(0,1)} $. For our computations it is enough to specify the holomorphic part only.
It is well known that the propagator in $\nu$-coordintates can be given in terms of almost holomorphic modular 
forms~\cite{Klemm:2015iya}, which for genus one read          
\begin{gleichung} 
\hat S^{\nu\nu}=\frac{c^2}{12}\left( E_2-\frac{3}{\pi {\rm Im}(\tau)}\right)=\frac{c^2}{12} \hat E_2\ 
\end{gleichung}\noindent
while for genus two curves with ${\rm Im}(\tau)_{pq}$, $p,q=1,2$ one obtains a very 
similar expression
\begin{gleichung} 
\hat S^{ij}=\frac{1}{2\pi i} \frac{C^i_p C^j_q}{10}\left(\frac{\partial}{\partial \tau_{pq}} 
\log(\chi_{10}(\tau))-\frac{5}{2 \pi} ({\rm Im}(\tau)^{-1})^{pq}\right) \ ,
\end{gleichung}\noindent
where now $\chi_{10}$ is the Igusa cusp form and the $C^i_p$ are constant invertible 
normalisation matrices.  For $\Sigma_{g>2}$ one can solve (\ref{equationsprop}) to get 
further interesting anholomorphic objects.   From the general structure one  can 
see that solving  (\ref{simplyfiedanomaly}) explicitly leads to $F^{(m,g)}$ 
that are polynomials of degree $3g+2n -3$ in the propagators with meromorphic 
but not an-holomorphic coefficients so that the total expression for $F^{(m,g)}$  is invariant 
on the monodromy group $\Gamma_{\Sigma_g}$ of  the family $\Sigma_g$. 
The meromorphic coefficients become simpler if one redefines the propagators themselves 
by meromorphic but not an-holomorphic forms, which we will do in section \ref{subsec_onedimholeq}.

\subsection{Fixing the Holomorphic Ambiguity}  
\label{subsection_Fha}
After integrating we still have a holomorphic ambiguity or recursion kernel in the free energies $F^{(n_1,n_2)}(\epsilon_1,\epsilon_2,\nu)$ or more specifically in the 
$F_n=F^{(n,0)}$.  Fixing this ambiguity can be done by imposing the so called gap condition at all conifold divisors with a suitable normalised 
vanishing cycle $\nu$, at which the leading behavior of each $F^{(n_1,n_2)}(\epsilon_1,\epsilon_2,\nu)$ can be determined by an expansion the refined BPS saturated Schwinger-Loop 
integral
\begin{eqnarray}  
\label{Schwinger} 
F(s,g_s,\nu)&=&\int_0^{\infty} \frac{d \sigma }{\sigma}\frac{\exp(-\sigma \nu)}{4\sinh(\sigma \epsilon_1/2)\sinh(\sigma\epsilon_2/2)} +\mathcal{O} (\nu^0) \\  
&=& \big{[}-\frac{1}{12}+\frac{1}{24} (\epsilon_1+\epsilon_2)^2 (\epsilon_1\epsilon_2)^{-1}\big{]}\log(\nu) \nonumber \nonumber \\ &&  
+ \frac{1}{\epsilon_1\epsilon_2} \sum_{g=0}^\infty \frac{(2g-3)!}{\nu^{2g-2}}\sum_{m=0}^g \hat B_{2g} \hat B_{2g-2m} \epsilon_1^{2g-2m} \epsilon_2^{2m} +\ldots \nonumber \\ &=& 
\big{[}-\frac{1}{12}+\frac{1}{24} s g_s^{-2}\big{]}\log(\nu) 
+ \big{[} -\frac{1}{240}g_s^2+\frac{7}{1440} s- 
\frac{7}{5760} s^2g_s^{-2} \big{]} \frac{1}{\nu^2} \nonumber \\ &&  
+ \big{[} \frac{1}{1008}g_s^4-\frac{41}{20160} s g_s^2 +\frac{31}{26880} s^2 -\frac{31}{161280} s^3 g_s^{-2}\big{]} \frac{1}{\nu^4}   +\mathcal{O} (\nu^0)  \nonumber\\ [ 3 mm] 
&& +  \,\,\mbox{contributions to $2(g+n)-2>4$}\, .  \nonumber 
\end{eqnarray} 
Here $g_s^2= (\epsilon_1 \epsilon_2)$, $s= (\epsilon_1 + \epsilon_2)^{2}$ and $\hat B_{m}=\left(\frac{1}{2^{m-1}}-1\right)\frac{B_m}{m!}$, where $B_m$ are the Bernoulli 
numbers defined by the generating function $t/(e^t-1)=\sum_{m=0}^\infty B_m \frac{t^m}{m!}$. This  determines the leading coefficient of the free 
energies and moreover implies that all other lower singular coefficients vanish. More precisely, the gap conditions, for example for the later defined sextic oscillator,  can be written as 
\begin{gleichung}
	F_n(\nu)	&= k_n	\frac{(1-2^{1-2n})(2n-3)!B_{2n}}{(2n)!}~\nu^{2-2n} + \mathcal O\left( \nu^0	\right)\quad \text{and}\\
	F_n^\pm(t_f^\pm)	&=	k_n^\pm	\frac{(1-2^{1-2n})(2n-3)!B_{2n}}{(2n)!}\cdot\left(-\frac{1}{2}\right)(-4)^n~\left(t_f^\pm\right)^{2-2n}		+	\mathcal O\left(\left(t_f^\pm\right)^0\right)
\label{gapcondition}
\end{gleichung}\noindent
where $n\geq2$. In \eqref{gapcondition} $\nu$ and $t_f^\pm$ label locally flat coordinates around the conifold loci associated to the vanishing cycles. 
In our model we could set the normalization constants $k_n$ and $k_n^\pm$ to unity. It is interesting that the upper gap condition is general in the sense 
that it is true for all anharmonic oscillators \eqref{eq:potential_class}, not only for the sextic. This is related to the fact that the gap condition is implied by the 
quantization condition of the underlying quantum mechanical system~\cite{Codesido:2016dld,ZinnJustin:2004ib,ZinnJustin:2004cg,Jentschura:2010zza}. The asymptotic behavior implied by the quantization condition reads
\begin{gleichung}
	-\int 	\log	\left[	\frac{\sqrt{2\pi}}{\Gamma\left( \nu+\frac{1}{2} \right)}	\right] ~\mathrm d\nu	&=	\frac{\nu^2}{2}\left(	\log \nu -\frac{3}{2}	\right)	-\frac{1}{24}\log \nu	-\frac{7}{\num{5760}}\frac{1}{\nu^2} \\
		&\qquad	+\frac{31}{\num{161280}}\frac{1}{\nu^4} 		-\frac{127}{\num{1290240}}\frac{1}{\nu^6}+\mathcal O\left(	\frac{1}{\nu^8}	\right)
\label{quant}
\end{gleichung}\noindent
and matches with the leading singular coefficient of the free energies. This universal singular behavior shows up for both examples we discuss in section \ref{sec_examples}.

The holomorphic ambiguity can be parametrized as a polynomial divided by the discriminant locus to the power $2n-2$. The degree of the numerator is bounded imposing regularity for $\xi\rightarrow\infty$. Actually, the degree can be further reduced by one as the highest degree monomial can be combined with appropriate lower degree ones to yield a constant contribution in $\xi$. This gives a constant contribution to the free energies which is unphysical. More precisely, this means
\begin{gleichung}
	h_n(\xi)		=	\frac{u_n(\xi)}{\left(	\prod_r\Delta_r^p	\right)^{2n-2}}~,
\label{parametrizationambi}
\end{gleichung}\noindent
where $u_n(\xi)$ is a polynomial whose degree is one less than that of the denominator. 
Requiring linear independence of the gap condition at all conifold loci fixes the holomorphic ambiguity completely.


\section{Examples}
\label{sec_examples}

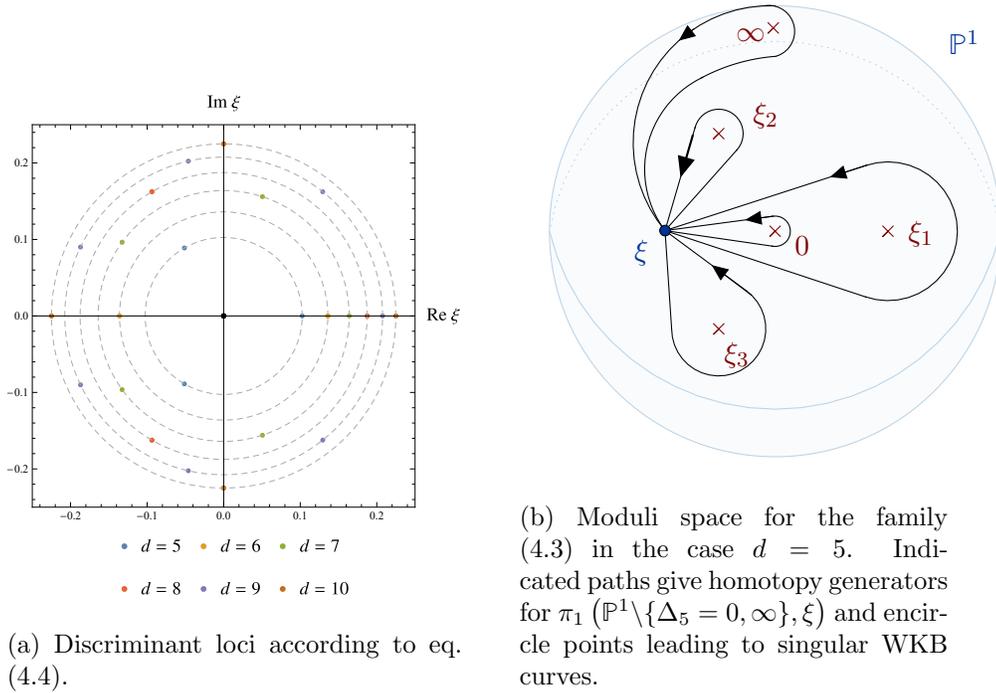
\begin{figure}
\centering
\begin{subfigure}[b]{0.425\textwidth}
        \includegraphics[width=\textwidth]{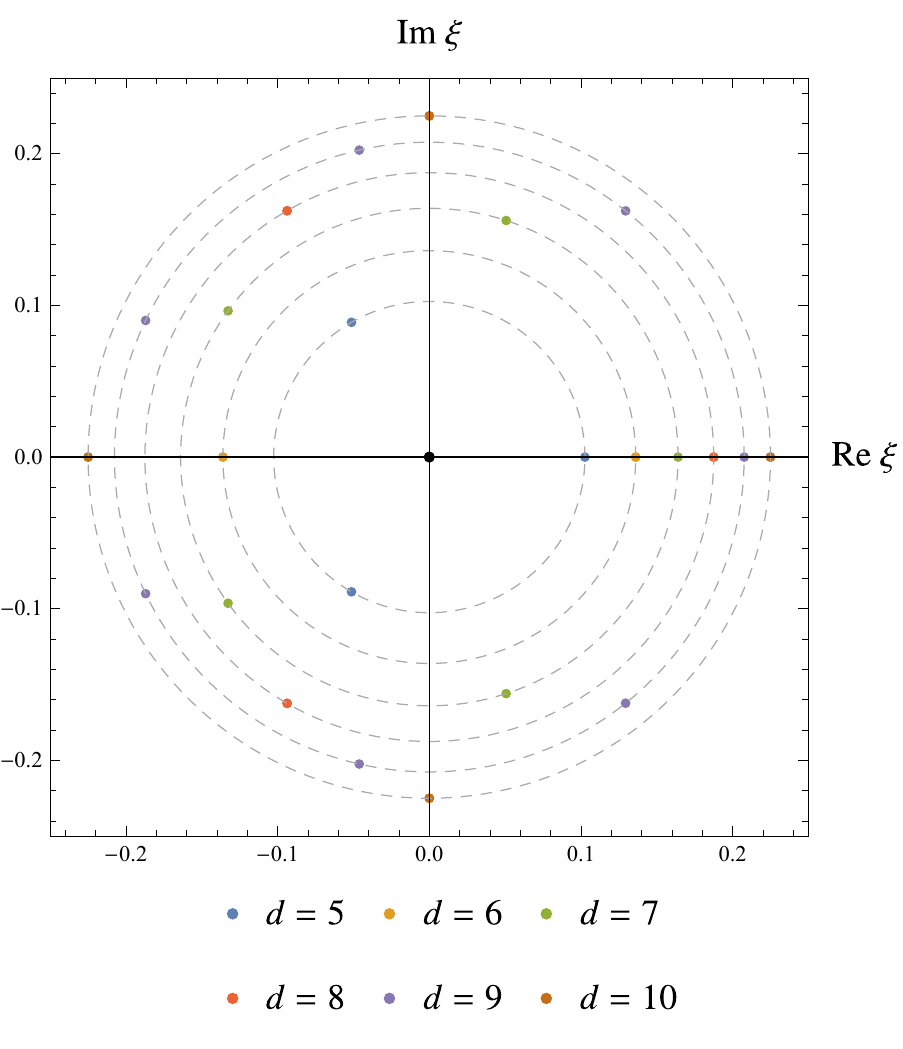}
        \caption{Discriminant loci according to eq. \eqref{eq:disc_loci_family}.}
        \label{fig:discriminant_loci}
\end{subfigure}      
\hspace*{0.5cm}
\begin{subfigure}[b]{0.4\textwidth}
\definecolor{qqqqff}{rgb}{0,0,1}
\definecolor{qqttzz}{rgb}{0,0.2,0.6}
\definecolor{bcduew}{rgb}{0.7372549019607844,0.8313725490196079,0.9019607843137255}
\definecolor{yqqqqq}{rgb}{0.5019607843137255,0,0}
\begin{tikzpicture}[line cap=round,line join=round,>=triangle 45,x=1cm,y=1cm]\clip(3.6018834093670242,0.5910064918071415) rectangle (10.515854046054992,7.3167944599561725);\draw [line width=0.4pt,color=bcduew,fill=bcduew,fill opacity=0.08] (7,4) circle (3cm);\draw [color=yqqqqq](6.3479564243418825,6.828913467447952) node[anchor=north west] {$\infty$};\draw [color=yqqqqq](7.12856601235504,4.0689009955443085) node[anchor=north west] {$0$};\draw [color=yqqqqq](8.63402736066613,4.277992849476403) node[anchor=north west] {$\xi_1$};\draw [color=yqqqqq](6.570987735202785,5.84618175396711) node[anchor=north west] {$\xi_2$};\draw [color=yqqqqq](6.187652669660609,2.6610158457348745) node[anchor=north west] {$\xi_3$};\draw [color=qqttzz](9.170696452425174,6.808004282054743) node[anchor=north west] {$\mathbb{P}^1$};\draw [shift={(6.25,5.299038105676658)},line width=0.4pt]  plot[domain=-0.7592738099005807:2.9235563834277123,variable=\t]({1*0.3246403175532128*cos(\t r)+0*0.3246403175532128*sin(\t r)},{0*0.3246403175532128*cos(\t r)+1*0.3246403175532128*sin(\t r)});\draw [line width=0.4pt] (6.492491370918076,5.083192626661436)-- (5.536083696357067,4.008717454797731);\draw [color=qqttzz](4.995829102247662,4.02708262475789) node[anchor=north west] {$\xi$};\draw [shift={(7,4)},line width=0.4pt]  plot[domain=-1.794242291474128:1.7545583615942644,variable=\t]({1*0.20183828154290853*cos(\t r)+0*0.20183828154290853*sin(\t r)},{0*0.20183828154290853*cos(\t r)+1*0.20183828154290853*sin(\t r)});\draw [shift={(8.5,4)},line width=0.4pt]  plot[domain=-1.9038449506530757:1.8913990797176676,variable=\t]({1*0.9228712550271748*cos(\t r)+0*0.9228712550271748*sin(\t r)},{0*0.9228712550271748*cos(\t r)+1*0.9228712550271748*sin(\t r)});\draw [shift={(6.25,2.700961894323342)},line width=0.4pt]  plot[domain=-3.1355648673304657:0.9533263402211793,variable=\t]({1*0.6258399740168744*cos(\t r)+0*0.6258399740168744*sin(\t r)},{0*0.6258399740168744*cos(\t r)+1*0.6258399740168744*sin(\t r)});\draw [line width=0.4pt] (5.938309273842235,5.3898138018088515)-- (5.536083696357067,4.008717454797731);\draw [line width=0.4pt] (8.215228512311695,4.877836290634665)-- (5.536083696357067,4.008717454797731);\draw [line width=0.4pt] (5.624171395648341,2.69718948757209)-- (5.536083696357067,4.008717454797731);\draw [line width=0.4pt] (6.62254462878062,3.2038397847143987)-- (5.536083696357067,4.008717454797731);\draw [line width=0.4pt] (6.97336284486576,4.200072871380773)-- (5.536083696357067,4.008717454797731);\draw [line width=0.4pt] (6.97098199612711,3.8002585487500746)-- (5.536083696357067,4.008717454797731);\draw [->,line width=0.4pt] (8.215228512311695,4.877836290634665) -- (7.727050094776595,4.719470436201506);\draw [line width=0.4pt] (8.210069709025397,3.1238540191664286)-- (5.536083696357067,4.008717454797731);\draw [->,line width=0.4pt] (6.97336284486576,4.200072871380773) -- (6.599052986929149,4.150238280187463);\draw [->,line width=0.8pt] (5.905222553817407,5.276206038606484) -- (5.74210743849574,4.71612805619285);\draw [->,line width=0.4pt] (6.6847911712228365,3.151108881988611) -- (6.155639077588703,3.5497351970363384);\draw [shift={(6.913797016979732,6.659427152930893)},line width=0.4pt]  plot[domain=-1.2867089737130133:1.7466741575887637,variable=\t]({1*0.3455466856910411*cos(\t r)+0*0.3455466856910411*sin(\t r)},{0*0.3455466856910411*cos(\t r)+1*0.3455466856910411*sin(\t r)});\draw [shift={(6.935871134109271,5.17752695375305)},line width=0.4pt]  plot[domain=1.5356230122111587:3.8373082536122562,variable=\t]({1*1.8236009749221642*cos(\t r)+0*1.8236009749221642*sin(\t r)},{0*1.8236009749221642*cos(\t r)+1*1.8236009749221642*sin(\t r)});\draw [shift={(7.197014731725328,3.829652604480532)},line width=0.4pt]  plot[domain=1.6452626768335061:2.0645046829200533,variable=\t]({1*2.5050203739592134*cos(\t r)+0*2.5050203739592134*sin(\t r)},{0*2.5050203739592134*cos(\t r)+1*2.5050203739592134*sin(\t r)});\draw [line width=1.2pt] (5.536083696357067,4.008717454797731)-- (5.536083696357067,4.008717454797731);\draw [shift={(6.6232375866629365,4.823356991193072)},line width=0.4pt]  plot[domain=2.014071255030222:3.7846663040756834,variable=\t]({1*1.3585069581955072*cos(\t r)+0*1.3585069581955072*sin(\t r)},{0*1.3585069581955072*cos(\t r)+1*1.3585069581955072*sin(\t r)});\draw [->,line width=0.4pt] (5.919691965870394,6.691759569230082) -- (5.71795078417024,6.534800152833271);\draw [shift={(7,4.71763654738694)},line width=0.4pt,color=bcduew]  plot[domain=3.3763925905682557:6.048385370201124,variable=\t]({1*3.084639721936007*cos(\t r)+0*3.084639721936007*sin(\t r)},{0*3.084639721936007*cos(\t r)+1*3.084639721936007*sin(\t r)});\draw [shift={(7,3.482610780807895)},line width=0.4pt,dotted,color=bcduew]  plot[domain=0.17078306840929616:2.970809585180497,variable=\t]({1*3.044288357586419*cos(\t r)+0*3.044288357586419*sin(\t r)},{0*3.044288357586419*cos(\t r)+1*3.044288357586419*sin(\t r)});\begin{scriptsize}\draw [color=yqqqqq] (7,4)-- ++(-2pt,-2pt) -- ++(4pt,4pt) ++(-4pt,0) -- ++(4pt,-4pt);\draw [color=yqqqqq] (8.5,4)-- ++(-2pt,-2pt) -- ++(4pt,4pt) ++(-4pt,0) -- ++(4pt,-4pt);\draw [color=yqqqqq] (6.25,5.299038105676658)-- ++(-2pt,-2pt) -- ++(4pt,4pt) ++(-4pt,0) -- ++(4pt,-4pt);\draw [color=yqqqqq] (6.25,2.700961894323342)-- ++(-2pt,-2pt) -- ++(4pt,4pt) ++(-4pt,0) -- ++(4pt,-4pt);\draw [fill=qqttzz] (5.536083696357067,4.008717454797731) circle (2pt);\draw [color=yqqqqq] (6.975287966822331,6.7049255896372575)-- ++(-2pt,-2pt) -- ++(4pt,4pt) ++(-4pt,0) -- ++(4pt,-4pt);
\end{scriptsize}
\end{tikzpicture}
\caption[Moduli space of the quintic family.]{Moduli space for the family \eqref{eq:riemsurf_class} in the case $d=5$. Indicated paths give homotopy generators for $\pi_1\left(\mathbb{P}^1\backslash \{\Delta_5 =0, \infty\},\xi\right)$ and encircle points leading to singular WKB curves.}
\label{fig:moduli_sphere_quintic}
\end{subfigure}
\caption[Moduli space of the WKB curve for anharmonic oscillators.]{Moduli space for the family \eqref{eq:riemsurf_class} in case of various degrees $d$.}
\label{fig:moduli_space_subfigures}
\end{figure}

Our two examples fall into the class of anharmonic oscillators  of the form
\begin{gleichung}
	V(x)	=	\frac{x^2}{2} - \mathfrak{g} x^d,
\label{eq:potential_class}
\end{gleichung}\noindent
which has also been studied in e.g. \cite{Codesido:2016dld,ZinnJustin:2004ib,ZinnJustin:2004cg,Jentschura:2010zza,Gahramanov:2015yxk,0305-4470-33-13-304}. In this work we focus on the cases where the corresponding WKB curve has genus two, namely the quintic ($d=5$) and symmetric sextic ($d=6$) case.
For any degree $d$ the coupling $\mathfrak{g}$ may be set to unity upon rescaling
\begin{equation}\label{eq:rescaling}
x \rightarrow x \cdot \mathfrak{g}^{1/(2-d)}, \quad y \rightarrow y\cdot \mathfrak{g}^{1/(2-d)}, \quad \xi \rightarrow \xi \cdot \mathfrak{g}^{2/(2-d)}
\end{equation}
so that $\xi$ remains as single modulus of the hyperelliptic curve
\begin{gleichung}
	\Sigma :\quad	y^2	=	2\xi	-x^2	+2x^d .
\label{eq:riemsurf_class}
\end{gleichung}\noindent
For potentials of the simple form \eqref{eq:potential_class} the critical values can be computed explicitly and for all $d$. Solving the common root condition one finds
\begin{equation}\label{eq:disc_loci_family}
\Delta_d(\xi) =  \xi\ \prod_{k=0}^{d-3} \left(\xi -  v \ \mathrm{e}^{2\pi \mathrm{i} \  \frac{2k}{d-2}} \right)\qquad \text{with} \quad v= \frac{d-2}{2}d^{\frac{d}{2-d}}>0
\end{equation}
such that the non-zero roots of the discriminant differ by roots of unity. Examples with different $d$ are shown in Fig. \ref{fig:discriminant_loci}, while Fig. \ref{fig:moduli_sphere_quintic} introduces homotopy generators relevant for monodromies using the example of the quintic.

It is clear that all WKB differentials are linear combinations of differentials $\frac{x^m}{y^k} \dd x,$ $ k=3n-1$. Their residues are given in appendix \ref{sec:app_res} for potentials of the form \eqref{eq:potential_class}. Especially we give necessary conditions on  $k, m$ and $d$ for non-zero residue. In all but one case these are not satisfied for the terms in $\mathcal{Q}_{2n}$ due to \eqref{eq:deg_x_pn} and \eqref{eq:wkb_Qn_polyn}. For $n\geq 1$ the residue of $\mathcal{Q}_{2n}$ at the point(s) at infinity vanishes, independent of $d>2$. Regarding the classical differential one finds
\begin{equation}\label{eq:wkb0_residue_even}
\mathrm{Res}_{\infty^\pm}\mathcal{Q}_0  =
\begin{cases}
0  						&\text{if } d\neq 6 \text{ is even } \ \text{ or } \ \ d \text{ is odd} \\
\frac{\pm 1}{2\sqrt{2}}	&\text{if } d=6 .
\end{cases}
\end{equation}
Furthermore, the WKB differentials $\mathcal{Q}_{2n}$ have no residues at the branch points. So except for the case\footnote{In this case the derivative $\partial_\xi y \ \dd x = y^{-1}\ \dd x$ is holomorphic.} $d=6,  n=0$ all $\mathcal{Q}_{2n}$ are differentials of the second kind and represent elements of $H^1(\Sigma,\mathbb{C})$.

\subsection{The Symmetric Sextic Oscillator}
\label{subsec_sextic}

\begin{figure}[h]
\centering
         \includegraphics[width=0.6\textwidth]{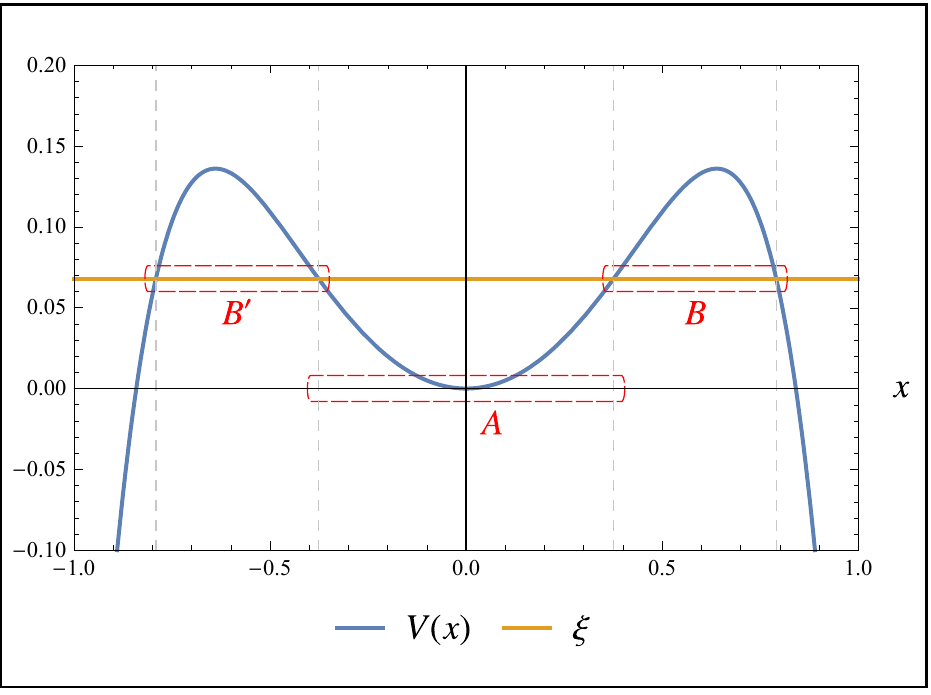}
        \caption{Sextic anharmonic potential with $A$- and $B$-cycle encircling pairs of turning points. The cycle $B'$ is the image of $B$ under $x\mapsto -x$.\\ }
        \label{fig_sextic1}
\end{figure}

\begin{figure}[h]
\centering
        \includegraphics[width=0.6\textwidth]{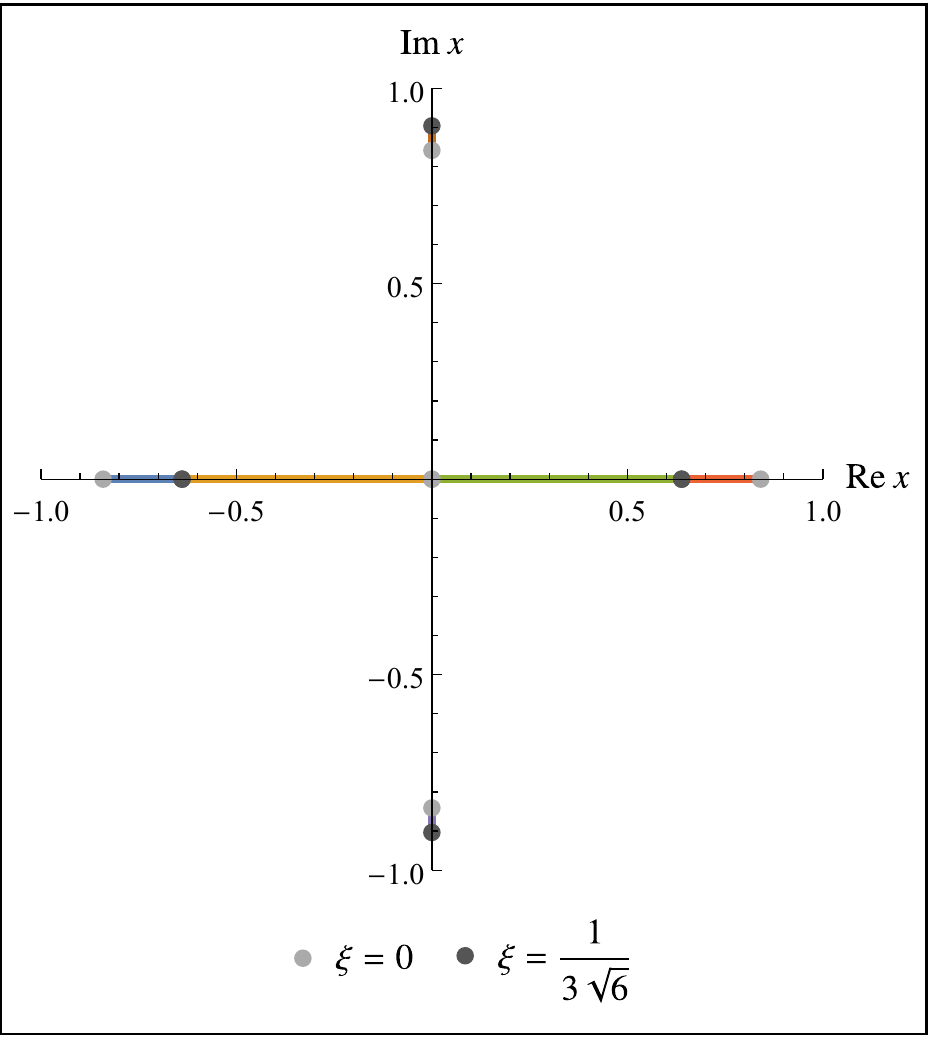}
        \caption{Trajectories of the turning points $x_i(\xi)$ as $\xi\in(0,\frac{1}{3\sqrt{6}})$ is varied between two zeroes of $\Delta_6$. Colors distinguish the six trajectories.}
        \label{fig_sextic2}
\end{figure}

 The corresponding WKB curve
\begin{gleichung}
	\Sigma^{(6)}:\quad	y^2	=	2\xi	-x^2	+2x^6 = \prod_{k=1}^{3} (x^2-x_k^2)
\label{defsextic}
\end{gleichung}\noindent
is of genus $g=2$. 
A plot of the potential as well as the homology cycles corresponding to pairs of real turning points are given in Fig. \ref{fig_sextic1} for generic small $\xi>0$. Over the complex numbers there are always six turning points, two of which are complex conjugated. This is illustrated in Fig. \ref{fig_sextic2} for $\xi$ varying between two non-negative roots of the (normalized) discriminant
\begin{gleichung}
	\Delta_6(\xi)	=	\xi\left(54\xi^2-1\right)^2.
\label{discriminant6}
\end{gleichung}\noindent

\subsubsection{Picard-Fuchs Operators and \QCO s}
To begin our discussion of the sextic oscillator we introduce quantum $A$- and $B$-periods $(\nu(\xi),\nu_D(\xi))$ as defined in \eqref{defaperiod} and \eqref{defbperiod}. The $A$- and $B$-cycle are defined in Fig. \ref{fig_sextic1} and encircle pairs of real branch points, see also Fig. \ref{fig_sextic2}. They are nomalized such that $\nu\hoch{0} = \xi + \hdots~$.
\noindent
The leading behavior of the classical periods can be determined as outlined in appendix \ref{sec_evperiodint}. Higher terms in the $\xi$-expansion are then generated using the Picard-Fuchs operator
\begin{gleichung}
	\mathcal L_\text{PF}\hoch 0	&=	\xi\left(54 \xi ^2-1\right)\ \partial_\xi^4  +2 \left(162 \xi ^2-1\right)\ \partial_\xi^3 + 354\xi~\partial_\xi^2 +30~\partial_\xi~.
\label{pfsextic0}
\end{gleichung}\noindent
Subsequently, quantum corrections to both periods are  obtained by applying diffe\-rential operators $\hbar^{2n}\mathcal D_{2n}$ to the classsical periods $(\nu\hoch 0 ,\nu_D\hoch 0)$, for example
\begin{gleichung}
	\mathcal D_2	&=	-\frac{45}{8}\xi\partial_\xi +\left(\frac{13}{45}-\frac{315}{8}\xi^2\right)\partial_\xi^2 -\frac{5}{16}\left(-1+54\xi^2\right)\xi\partial_\xi^3 \\
	\mathcal D_4	&=	\frac{7 + \num{5670}\xi^2}{192\xi-\num{10368}\xi^3}\partial_\xi	+ \frac{\num{2033}+\num{247050}\xi^2}{960-\num{51840}\xi^2}\partial_\xi^2	+\frac{7-\num{5724}\xi^2-\num{364500}\xi^4}{\num{2880}\xi(-1+54\xi^2)}\partial_\xi^3~.
\label{qcosextic}
\end{gleichung}\noindent
%
 This leads to the following expansions for the first few WKB orders of the $A$-period 
\begin{gleichung}
	\nu\hoch{0}(\xi) &= \xi + \frac{5 }{2}\xi^3+ \frac{693 }{16}\xi^5+ \frac{\num{36465 }}{32}\xi^7 + \mathcal O(\xi^9)\\
	\nu\hoch{2}(\xi) &=	\frac{25 }{8}\xi +\frac{\num{5145}}{32} \xi^3+	\frac{\num{1096095 }}{128}\xi^5+ \mathcal O(\xi^7)\\
	\nu\hoch{4}(\xi) &=	\frac{\num{21777} }{256}\xi   +\frac{\num{8703695 }}{512}\xi^3+ \frac{\num{16457464023 }}{\num{8192}}\xi^5 + \mathcal O(\xi^7)\\
	\nu\hoch{6}(\xi) &=	\frac{\num{12746305} }{\num{2048}}\xi  +\frac{\num{49058686105 }}{\num{16384}}\xi^3+\frac{\num{42777946276065 }}{\num{65536}}\xi^5+  \mathcal O(\xi^7)~,
\label{pertperiods}
\end{gleichung}\noindent
respectively the $B$-period
\begin{gleichung}
	\nu_D\hoch 0 (\xi)	 = &\frac{\pi }{4 \sqrt{2}}-\left(1-\log \left(\frac{\xi}{2^{3/2}} \right)\right)\xi	+\left(\frac{17}{3}+\frac{5 }{2}\log \left(\frac{\xi}{2^{3/2}} \right) \right) \xi^3 + \mathcal O(\xi^5)\\
	\nu_D\hoch 2 (\xi)	 = &	-\frac{1}{24}\frac{1}{\xi}+\left(\frac{149}{16}+\frac{25}{8}\log \left(\frac{\xi}{2^{3/2}} \right)\right)\xi	+	\left(\frac{\num{62223}}{128} 	+	\frac{\num{5145}}{32}\log \left(\frac{\xi}{2^{3/2}} \right)\right)\xi^3+ \mathcal O(\xi^5)\\
	\nu_D\hoch 4 (\xi)	 = &	\frac{7}{\num{2880}}\frac{1}{\xi^3}+\frac{43}{384}\frac{1}{\xi}	+	\left(\frac{\num{1724749}}{\num{5120}} +   \frac{\num{21777}}{256}\log \left(\frac{\xi}{2^{3/2}} \right)	\right)\xi \\
		& \hspace{2.7cm}	+	\left(	\frac{\num{366995597}}{\num{6144}}+ \frac{\num{8703695}}{512}\log \left(\frac{\xi}{2^{3/2}} \right)	\right)\xi^3		+ \mathcal O(\xi^5)\\
	\nu_D\hoch 6 (\xi)	 = &	-\frac{31}{\num{40320}}\frac{1}{\xi^5}-\frac{425}{\num{32256}}\frac{1}{\xi^3}-\frac{\num{32793}}{\num{14336}}\frac{1}{\xi}+	
	\left( \frac{\num{4824402623}}{\num{86016}}+\frac{\num{12746305}}{\num{1024}}  \log \left(\frac{\xi}{2^{3/2}} \right)	\right)\xi 	\\
	& \hspace{2cm}+	
	\left(	\frac{\num{31906379165357 }}{\num{2752512}}+\frac{\num{49058686105}}{\num{16384}}\log \left(\frac{\xi}{2^{3/2}} \right)	\right)\xi^3		+ \mathcal O(\xi^5)~. 
\label{nonpertperiods}
\end{gleichung}\noindent

At any given order in $\hbar^2$ the $A$- and $B$-period are annihilated by a corresponding Picard-Fuchs operator which we collect in appendix \ref{sec_highpfops}. 
%
\noindent


\subsubsection{Quantum Free Energies from Quantum Mechanics}
As explained in subsection \ref{subsec:connection} we can construct  quantum free energies in the WKB framework. This is done by firstly computing the quantum periods \eqref{pertperiods} and \eqref{nonpertperiods}, secondly using the inverse of the $A$-period as the mirror map to express the $B$-period in terms of $\nu$ and thirdly integrating with respect to $\nu$. Thus, calculating free energies up to $F_n$ requires the computation of the $A$- and $B$-period up to order $\hbar^{2n}$. %
We find
\begin{gleichung}
	F_0(\nu)	&=	\frac{\nu ^2}{2}  \left( \log \left(\frac{\nu}{2^{3/2}} \right)-\frac{3}{2}	\right)	+\frac{\pi   }{4 \sqrt{2}}\nu	+\frac{17 }{12}	\nu ^4+\frac{\num{10727}}{960} \nu ^6 +\frac{\num{55747}}{336}  \nu ^8+	\mathcal O(\nu^{10})\\
F_1(\nu)	&=	-\frac{1}{24}\log \nu 		+\frac{221}{48} \nu ^2	+\frac{\num{38459} }{384}\nu ^4+\frac{\num{3442219}}{\num{1152}} \nu ^6		+\frac{\num{2484506287}}{\num{24576}}	 \nu ^8+\mathcal O(\nu^{10})\\
	F_2(\nu)	&=	-\frac{7}{\num{5760}}\frac{1}{\nu^2}	+\frac{\num{2283899}}{\num{15360}} \nu ^2		+\frac{\num{14725045}}{\num{1152}} \nu ^4	+\frac{\num{2619604188383 }}{\num{2949120}}\nu ^6\\
		&\hspace{8cm}+\frac{\num{284059718609}}{\num{5120}} \nu ^8+\mathcal O(\nu^{10})\\
	F_3(\nu)	&=	\frac{31}{\num{161280} \nu^4}	+\frac{\num{1642757413} \nu ^2}{\num{129024}}	+\frac{\num{3084767116889} \nu ^4}{\num{1179648}} +\frac{\num{334449226613647} \nu ^6}{\num{983040}}\\
		&\hspace{6.5cm}	+\frac{\num{191100242149408097} \nu ^8}{\num{5505024}}+\mathcal O(\nu^{10})~.
\label{sexticfn}
\end{gleichung}\noindent
These expressions can be compared with the topological string computation which we will be presented shortly. As initial datum we use the classical free energy $F_0(\nu)$. Moreover, the first quantum correction $F_1$ will be used to fix the parameters in the ansatz \eqref{ansatzfreeenergies}.


\subsubsection{Solving the Holomorphic Anomaly Equation for the Reduced Sextic}
\label{subsec_onedimholeq} 

So far, we have used geometry within the WKB method to calculate quantum periods. On the other hand using the claim stated in subsection \ref{subsec:connection} we can use string theoretical methods, namely the holomorphic anomaly equation, to compute free energies related to quantum periods by \eqref{deffreeenergies}. In this paragraph we make this explicit by solving the holomorphic anomaly equation for the sextic curve \eqref{defsextic}. Our approach to solve this equation follows the procedure in \cite{Haghighat:2008gw}.

Starting with the classical free energy $F_0$ \eqref{sexticfn} we can compute the Yukawa coupling \eqref{yukawa} in the coordinate $\xi$ as
\begin{gleichung}
	C_{\xi\xi\xi}	=	\frac{1}{\xi(1-54\xi^2)}~.
\label{yukawa}
\end{gleichung}\noindent
By comparison with the quantum mechanical computations \eqref{sexticfn} we can fix the parameters in \eqref{ansatzfreeenergies} and find for $F_1$ 
\begin{gleichung}
	F_1	=	-\frac{1}{24} \log\left(		\xi\left(1-54\xi^2\right)^2	\right)~.
\label{f1}
\end{gleichung}\noindent
We can now compute the propagator from \eqref{equationsprop} imposing $A_\xi=0$ and obtain
\begin{gleichung}
	S^{\xi\xi}	=	-15\xi^2 + \frac{225}{4}\xi^4 + \frac{6975}{4}\xi^6	+	\frac{8253225}{128}\xi^8 + \mathcal O\left(\xi^{10}\right)~.
\label{propagator}
\end{gleichung}\noindent
As it turns out, the parameters $a$ and $b$ in the ansatz \eqref{ansatzfreeenergies} for $\mathcal F^{(0,1)}$ may be set to zero as they will only affect the constant terms of the free energies which in our case are unphysical since the WKB method only determines the derivative of the free energies. Then the Christoffel symbols and the covariant derivative of the propagator are given by
\begin{gleichung}
	\Gamma_{\xi\xi}^\xi 	&=	- C_{\xi\xi\xi}S^{\xi\xi}\qquad\quad\text{and} \\
	D_\xi S^{\xi\xi}	&=	-C_{\xi\xi\xi}S^{\xi\xi}S^{\xi\xi} -30\xi~.
\label{chriscovariant}
\end{gleichung}\noindent
By considering NS free energies the first part in \eqref{simplyfiedanomaly} drops out. After writing everything as a polynomial in the propagator with coefficients being rational functions in $\xi$ we obtain
\begin{gleichung}
	F_2	=	\frac{\left(	1-270\xi^2	\right)^2}{1152\xi^2\left(	1-54\xi^2	\right)^2}~S^{\xi\xi}	+	h_2(\xi)~.
\label{f2}
\end{gleichung}\noindent
The holomorphic ambiguity $h_2(\xi)$ can be fixed by the gap condition at all three conifold loci
\begin{gleichung}
	\Delta =	\xi\left(	1-54\xi^2	\right)^2	= 0	\quad\Rightarrow\quad \xi_c \in \{0,\pm\frac{1}{\sqrt{54}}\}~.
\label{conifoldloci}
\end{gleichung}\noindent
The propagator is transformed to a different conifold locus with
\begin{gleichung}
	S_f^{\xi\xi}		=	\frac{2}{C_{\xi\xi\xi}}\partial_\xi \mathcal F^{(0,1)}_f	=	\frac{1}{C_{\xi\xi\xi}}\partial_\xi \log\left( \frac{\partial \xi}{\partial t_f}	\right)~,
\label{propconifold}
\end{gleichung}\noindent
where $t_f$ is an appropriate flat coordinate at the conifold loci. Imposing the gap condition at these three points is enough to fix the ambiguity completely. If we assume that the gap conditions give linearly independent conditions we will obtain $6(n-1)$ conditions, which equals the number of parameters in the ambiguity \eqref{parametrizationambi}. The result for $h_2$ reads
\begin{gleichung}
	h_2(\xi)	=	\frac{-7+\num{3924}\xi^2+\num{461700}\xi^4}{\num{5760}\xi^2\left(	1-54\xi^2	\right)^2}~.
\label{ambif2}
\end{gleichung}\noindent
Expanding at the conifold locus $\xi_c=0$ in the flat coordinate $\nu$ we find\footnote{Here and in the following we have dropped the unphysical constant term.}
\begin{gleichung}
	F_2	=	\frac{7}{\num{5760}}\frac{1}{\nu^2}	+	\frac{\num{2283899}}{\num{15360}}\nu^2	+	\frac{\num{14725045}}{\num{1152}}\nu^4	+	\frac{\num{2619604188383}}{\num{2949120}}\nu^6 	+	\mathcal O\left(	\nu^8	\right)~.
\label{f2full}
\end{gleichung}\noindent
For the higher free energies we can now go on and compute them recursively. For $F_3$ we find
\begin{small}
\begin{gleichung}
	F_3	&=	\frac{-1+270\xi^2}{\num{414720}\xi^4\left( 1-54\xi^2\right)^4}\left(	42 - \num{7254}\xi +\num{5580252}\xi^4 +\num{103663800}\xi^6 + \num{1771470000}\xi^8\right.	\\
	&\qquad \left.	-15S^{\xi\xi}	 + \num{2430}\xi^2S^{\xi\xi} +\num{218700}\xi^4S^{\xi\xi} + \num{59049000}\xi^6S^{\xi\xi}	+	5\left(S^{\xi\xi}\right)^2 -\num{2700}\xi^2\left(S^{\xi\xi}\right)^2\right.	\\
	&\qquad \left. + \num{364500}\xi^4\left(S^{\xi\xi}\right)^2	\right)	\\
	&\qquad	+\frac{31 -\num{7251}\xi^2 +\num{3442320}\xi^4 + \num{1396047960}\xi^6 + \num{44918355600}\xi^8 + \num{296780274000}\xi^{10}}{\num{161280} \xi^4 \left(1-54\xi^2\right)^4}~.
\label{f3}
\end{gleichung}\noindent
\end{small}\noindent
Expressing higher free energies at the conifold locus $\xi_c=0$ as well in terms of the locally flat coordinate $\nu$ we recover quantum mechanical results.

We checked with explicit computations in the WKB framework as well as in the topological string framework that up to order $\hbar^{10}$ the free energies agree. This provides a constructive confirmation of the conjecture proposed in \cite{Codesido:2016dld}, at least with regard to the current example. Moreover, the gap condition at all conifold loci gives enough information to fix the holomorphic ambiguities completely.

From a computational point of view solving the holomorphic anomaly equation is more involved. The geometrical methods used to compute quantum periods are more efficient. In particular, using quantum differential operators $\mathcal{D}_{2n}$ simplifies and accelerates the calculation of the quantum periods enormously. With a modern computer quantum differential operators can be computed quickly.


\subsubsection{Reduction to the Elliptic Case}
A special property of the symmetric sextic potential is that all WKB periods along $A, B$ and $B'$ reduce to elliptic ones. 

To see this, first consider the Klein four-group $\mathbb{Z}_2 \times \mathbb{Z}_2 \subset \operatorname{Aut}(\Sigma^{(6)})$ of holomorphic automorphisms generated by the hyperelliptic involution $i_1 : (y,x)\mapsto (-y,x)$ and the reflection $i_2 : (y,x)\mapsto (y,-x)$. The  involution $i_3= i_2 i_1$ simultaneously reversing position and momentum has no fixed points, as for $\xi \neq 0$ none of the branch points $(0,x_k)$ equals $(0,0)$. Hence it defines an unramified two-sheeted covering\footnote{This makes use of the following theorem \cite{bobenko2011computational}: Let $\mathcal{R}$ be a (compact) Riemann surface and $G$ be a finite group of holomorphic automorphisms of order $|G|$. Then $\mathcal{R}/G$ is a Riemann surface with the complex structure determined by the condition that the canonical projection $\pi : \mathcal{R} \rightarrow \mathcal{R}/G$ is holomorphic. This is a $|G|$-sheeted covering, ramified at the fixed points of $G$.}
\begin{equation}
c : 
\begin{cases} 
   \Sigma^{(6)}& \rightarrow  \qquad \Sigma^{(6)}/i_3 \\
   (y,x)      & \mapsto (Y,X)=(yx,x^2)
  \end{cases}
\end{equation}
mapping to the elliptic curve $\mathcal{E} =\Sigma^{(6)} /i_3$ given by
\begin{gleichung}
	\mathcal{E}:	Y^2	=	X \prod_{i=k}^{3} (X-x_k^2)~.
\label{defelliptic}
\end{gleichung}\noindent
Even though $\Sigma^{(6)}$ may also be regarded as double covering of the elliptic curve $ \Sigma^{(6)}/i_2 : y^2 = 2\xi - X + 2 X^3$, this is not the correct geometry for the WKB periods: na\"{i}ve substitution in the classical differential $y \ \dd x$ leads to the tentative one-form
\begin{equation}
\sqrt{2\xi - X +2X^3} \frac{\dd X}{2\sqrt{X}}~,
\end{equation}
which however is multi-valued in any sheet of $\Sigma^{(6)}/i_2$ (note that $(y,X=0)$ is not a branch point). Expanding the fraction by $\sqrt{X}$, we obtain the one-form $\omega = Y \dd X / 2X$ well-defined on $\Sigma^{(6)}/i_3$. 

By mathematical induction it can then be shown that all WKB differentials $\mathcal{Q}_n=  Q_n(x) \ \dd x$ may be written as $\mathcal{Q}_n = p_n(x) \ \dd x / y^{3n-1}$ with polynomials $p_n(x)=(-1)^n p_n(-x)$ of well-defined parity (given that the potential $V(x)$ is an even polynomial). Thus, they are invariant under the action of $i_3$,
\begin{equation}
{i_3}^* \ \mathcal{Q}_n = \frac{p_n(-x) \ \dd (-x)}{(-y)^{3n-1}} = \mathcal{Q}_n
\end{equation}
and there exist polynomials $\tilde{p_n}(X)$ such that all WKB differentials become pullbacks of meromorhphic forms on $\mathcal{E}$,
\begin{equation}
\mathcal{Q}_n =  c^* \left( \frac{\tilde{p_n}(X) \ \dd X}{Y^{3n-1}} \right)~.
\end{equation}
Note that the holomorphic one-form $Y^{-1} \ \dd X= \partial_\xi \ \omega$ on $\mathcal{E}$ corresponds to its pullback $y^{-1} \ \dd x = \partial_\xi  \ (y \ \dd x)$ on $\Sigma$.

As the space $H_1(\mathcal{E},\mathbb{C})$ has smaller dimension than $H_1(\Sigma,\mathbb{C})$, we need to check that the classical $A$- and $B$-periods on the sextic (and thus their quantum counterparts) map to periods on $\mathcal{E}$. One way to verify this is to note that the Picard-Fuchs operator for the one-form $\omega$ reads
\begin{equation}
\mathcal{L}^{(0)}_\mathcal{E} = \xi(54\xi^2-1) \ \partial_\xi^3 + (162\xi^2 - 1) \ \partial_\xi^2 + 30\xi \ \partial_\xi
\end{equation}
and annihilates
\begin{align}
\nu^{(0)}(\xi) &= \xi  \ \, _3F_2\left(\frac{1}{6},\frac{1}{2},\frac{5}{6};1,\frac{3}{2};54 \xi ^2\right) \nonumber \\
			 &= \xi +\frac{5 \xi ^3}{2}+\frac{693 \xi ^5}{16}+\frac{36465 \xi ^7}{32}+\frac{37182145 \xi ^9}{1024}+\mathcal{O}\left(\xi ^{11}\right)
\end{align}
as well as	 
\begin{align}
\nu_D^{(0)}(\xi) &= \frac{\pi }{2^{5/2}}-\frac{\pi}{2}  \  \xi \   G_{3,3}^{2,1}\left(54 \xi ^2\Bigg|
\begin{array}{c}
 \frac{1}{2},\frac{1}{6},\frac{5}{6} \\
 0,0,-\frac{1}{2} \nonumber \\
\end{array}
\right) \\
			&= \frac{\pi }{4 \sqrt{2}} + \log\left(\frac{\xi}{2^{3/2}}\right)\left( \xi +\frac{5 \xi ^3}{2}+\frac{693 \xi ^5}{16}+\frac{36465 \xi ^7}{32}+\frac{37182145 \xi ^9}{1024}+\mathcal{O}\left(\xi ^{11}\right) \right) \nonumber \\
				& \quad  -\xi+\frac{17 \xi ^3}{3}+\frac{18027 \xi ^5}{160}+\frac{1394891 \xi ^7}{448}+\frac{3751204337 \xi ^9}{36864} +O\left(\xi ^{11}\right).
\end{align}
This leads precisely to the subsystem spanned by the classical periods in \eqref{pertperiods} and \eqref{nonpertperiods} and the residue  of $(y\ \dd x, \Sigma)$ or $(\omega, \mathcal{E})$. One can also check these findings by explicit computation of periods on $\mathcal{E}$.


\subsection{The Quintic Oscillator}
\label{subsec_quintic}

\begin{figure}[h]
\centering
          \includegraphics[width=0.6\textwidth]{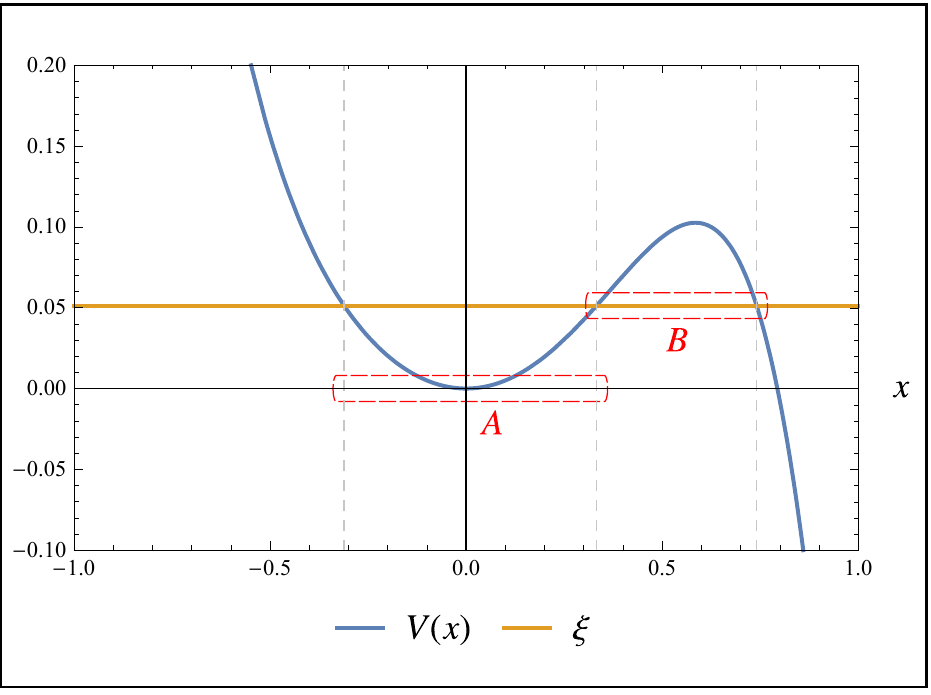}
        \caption{Quintic anharmonic potential with $A$- and $B$-cycle encircling pairs of turning points.}
        \label{fig:quintic_potential_a}
\end{figure}

\begin{figure}[h]
\centering
        \includegraphics[width=0.6\textwidth]{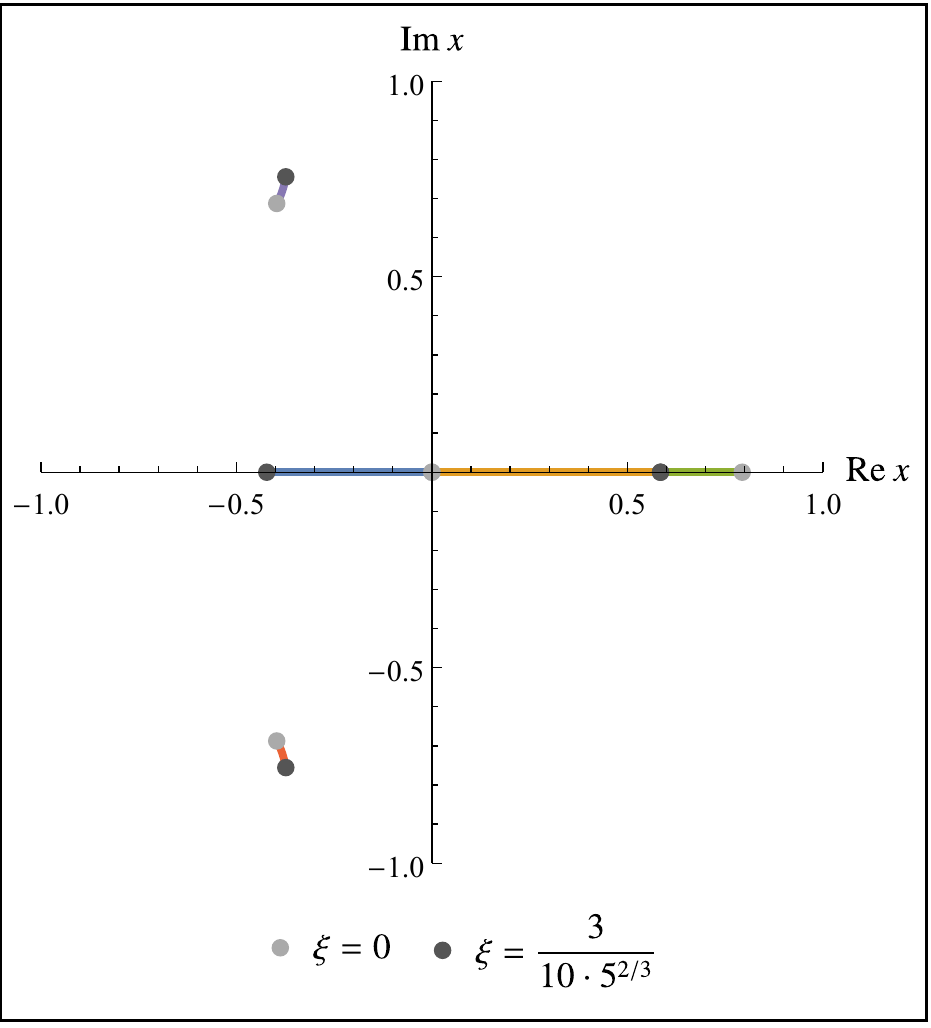}
        \caption{Trajectories of the turning points $x_i(\xi)$ as $\xi\in\left(0,\frac{3}{10\cdot5^{2/3}}\right)$ is varied between two zeroes of $\Delta_5$. Colors distinguish the five trajectories.}
        \label{fig:quintic_potential_b}
\end{figure}

We now turn to the quintic potential, $d=5$, which leads to the  family of genus-two curves
\begin{equation}
	\Sigma_{(5)} : \quad y^2 = 2\xi - x^2 + 2 x^5~.
\label{defquintic}
\end{equation}
The quintic curve can not be regarded as multi-cover of an elliptic curve and is in this sense more generic. The potential and the homology cycles corresponding to pairs of real turning points are given in Fig. \ref{fig:quintic_potential_a} for generic small $\xi>0$. Moreover, in Fig. \ref{fig:quintic_potential_b} the movement of the branch points is visualized. There are three real branch points and one pair of complex conjugated branch points. Additionally one branch point is at infinity. The quintic curve gets singular if the moduli $\xi$ is tuned to a root of the normalized discriminant
\begin{equation}
	\Delta_5(\xi) =  \xi \ (25000\xi^3 - 27)~.
\label{eq:quintic_discriminant}
\end{equation}

\subsubsection{Picard-Fuchs Operators and \QCO s}
The quantum periods $\nu$ and $\nu_D$ are defined by equation \eqref{defaperiod} and \eqref{defbperiod}, with the $A$- and $B$-cycle encircling branch points as shown in Fig. \ref{fig:quintic_potential_a} and \ref{fig:quintic_potential_b}. Classical periods can be determined from the Picard-Fuchs operator
\begin{gleichung} \label{eq:pfe_classical_quintic}
	\mathcal{L}^{(0)}_{\mathrm{PF}} =& \left( \xi (25000\xi^3 - 27)  \ \partial_\xi^4 + (6400000 \xi^3 - 1728) \ \partial_\xi^3 +  10160000 \xi^2 \ \partial_\xi^2 \right. \\
								&\quad\left. +  1120000 \xi \ \partial_\xi  + 104 720 \right),
\end{gleichung}\noindent
together with the leading behavior of the classical periods, which is determined in appendix \ref{sec_evperiodint}. Quantum corrections to the classical periods are easily obtained by applying differential operators $\hbar^{2n}\mathcal D_{2n}$ to the classical periods $(\nu\hoch 0,\nu_D\hoch 0)$, for example 
\begin{gleichung}
	\mathcal{D}_{2} &= -\frac{5}{567} \left(25000 \xi^3-27\right) \xi \partial_\xi^3 -\left(	\frac{11}{56}-\frac{137500}{189}\xi^3	\right) \partial_\xi^2 - \frac{\num{8750}}{81} \xi^2 \partial_\xi^1 -\frac{1375}{81} \xi \partial_\xi^0 \\
	\mathcal{D}_{4} &= -\frac{\left(39500000000 \xi^6+89640000 \xi^3-5103\right) }{77760 \xi \left(25000 \xi^3-27\right)}\partial_\xi^3-\frac{125 \xi \left(31900000 \xi^3+37557\right)}{2592 \left(25000 \xi^3-27\right)}\partial_\xi^2  \\ 
				&\quad -\frac{5 \left(172100000 \xi^3+111807\right) }{3888 \left(25000 \xi^3-27\right)}\partial_\xi^1-\frac{9163 \left(200000 \xi^3+27\right) }{62208 \xi \left(25000 \xi^3-27\right)}\partial_\xi^0~.
\label{diffop_quintic}
\end{gleichung}\noindent
The first few WKB orders of the quantum $A$-period read
\begin{gleichung*}
	\nu\hoch 0(\xi) 	&=	\xi +\frac{315}{16}\xi^4 + \frac{\num{692835}}{128}\xi ^7 + \frac{\num{9704539845}}{\num{4096}}\xi ^{10} + \mathcal O(\xi^{13})	\\
	\nu\hoch 2(\xi) 	&=	\frac{\num{1085}}{32}\xi ^2 + \frac{\num{15570555}}{512}\xi ^5 + \frac{\num{456782651325}}{\num{16384}}\xi ^8 + \frac{\num{6734319857340075}}{\num{262144}}\xi ^{11} + \mathcal O(\xi^{14})	\\
	\nu\hoch 4(\xi) 	&=	\frac{\num{1107}}{256} + \frac{\num{96201105}}{\num{2048}}\xi ^3 + \frac{\num{4140194663605}}{\num{32768}}\xi ^6 + \frac{\num{489884540580510075}}{\num{2097152}}\xi ^9 + \mathcal O(\xi^{12}) 
\end{gleichung*}\noindent
\begin{gleichung}
	\nu\hoch 6(\xi) 	&=	\frac{\num{118165905}}{\num{8192}}\xi + \frac{\num{30926063193025}}{\num{131072}}\xi ^4 + \frac{\num{2364285204614844225}}{\num{2097152}}\xi ^7 \\
	&	\hspace{5.5cm} + \frac{\num{223004451972549877775145}}{\num{67108864}}\xi^{10} + \mathcal O(\xi^{13})
\label{eq:Aperiod_hbar0_quintic}
\end{gleichung}\noindent
and respectively for the $B$-period
\begin{gleichung}
	\nu_D\hoch 0(\xi) 	&=	\frac{\sqrt{\pi }\ \Gamma \left(\frac{5}{3}\right)}{2\cdot 2^{2/3}\ \Gamma \left(\frac{13}{6}\right)} - \left( 1 - \log \left(\frac{\xi}{2^{5/3}} \right)	\right)\xi 	-\frac{7 \pi\  \Gamma \left(-\frac{1}{3}\right)}{9\ \Gamma \left(-\frac{1}{6}\right) \Gamma \left(\frac{5}{6}\right)}\xi^2 \\
					&\hspace{6cm}	+\frac{935 \sqrt{\pi }\ \Gamma \left(\frac{2}{3}\right)}{324\cdot 2^{2/3}\ \Gamma \left(\frac{7}{6}\right)}\xi^3 + \mathcal O(\xi^4)	\\
	\nu_D\hoch 2(\xi) 	&=	-\frac{1}{24}\frac{1}{\xi}	-\frac{11 \sqrt{\pi }\ \Gamma \left(\frac{1}{3}\right)}{72\cdot2^{1/3}\ \Gamma \left(\frac{5}{6}\right)}	+	\frac{385 \sqrt{\pi }\ \Gamma \left(\frac{2}{3}\right)}{24\cdot 2^{2/3}\ \Gamma \left(\frac{1}{6}\right)}\xi 	+ \left(	\frac{\num{66595}}{576} \right. \\
					&\hspace{6cm}\left.+ \frac{\num{1085}}{32}\log \left(\frac{\xi}{2^{5/3}} \right) 	\right)\xi^2	 + \mathcal O(\xi^3)	\\
	\nu_D\hoch 4(\xi) 	&=	\frac{7}{\num{2880}}\frac{1}{\xi^3} 	+	\left(	 \frac{\num{16853}}{768} + \frac{\num{1107}}{256}\log \left(\frac{\xi}{2^{5/3}} \right)	\right)	-	\frac{\num{4161703} \sqrt{\pi }\ \Gamma \left(\frac{1}{3}\right)}{\num{82944}\cdot 2^{1/3}\ \Gamma \left(\frac{5}{6}\right)}\xi \\
	&\hspace{6cm}	- \frac{\num{450756} \cdot 2^{1/3} \sqrt{\pi }\ \Gamma \left(-\frac{7}{3}\right)}{\num{150643225}\ \Gamma \left(-\frac{59}{6}\right)}\xi^2 + \mathcal O(\xi^3)	\\
	\nu_D\hoch 6(\xi) 	&=	-\frac{31}{\num{40320}}\frac{1}{\xi^5}+\frac{53}{\num{6144}}\frac{1}{\xi^2}	+\frac{\num{1641726} \cdot 2^{1/3} \sqrt{\pi }\ \Gamma \left(-\frac{7}{3}\right)}{\num{68542667375}\ \Gamma \left(-\frac{65}{6}\right)}   + \left(	\frac{\num{8031474795}}{\num{114688}} \right. \\
					&\hspace{1cm} \left.+ \frac{\num{118165905}}{\num{8192}}\log \left(\frac{\xi}{2^{5/3}} \right)	 	\right)\xi +	 \frac{\num{18570744}\cdot 2^{2/3} \sqrt{\pi }\ \Gamma \left(-\frac{11}{3}\right)}{\num{490839606425}\ \Gamma \left(-\frac{85}{6}\right)}\xi^2	+\mathcal O(\xi^3)	~.
\label{nonperturbativequintic}
\end{gleichung}\noindent
It is interesting that transcendental numbers appear in the $B$-period. Comparing with the fundamental system\footnote{The subscript indicates the corresponding root of the indicial equation.} of the Picard-Fuchs operator for the classical periods 
\begin{gleichung*}
	\Pi_0(\xi)	& = \, _4F_3\left(-\frac{7}{30},-\frac{1}{30},\frac{11}{30},\frac{17}{30};\frac{1}{3},\frac{2}{3},\frac{2}{3};\frac{25000 \xi ^3}{27}\right) \\
			& = 1+\frac{6545 \xi ^3}{648}+\frac{1682469481 \xi ^6}{839808}+ \mathcal{O} \left(\xi ^{9}\right)  \\  \\
	\Pi_1(\xi)	& = \xi \  \, _4F_3\left(\frac{1}{10},\frac{3}{10},\frac{7}{10},\frac{9}{10};\frac{2}{3},1,\frac{4}{3};\frac{25000 \xi ^3}{27}\right) \\
			& =\xi +\frac{315 \xi ^4}{16}+\frac{692835 \xi ^7}{128}+\mathcal{O}\left(\xi ^{10}\right) 
\end{gleichung*}\noindent
\begin{gleichung}
	\Pi_2(\xi)	& = \xi ^2 \ \, _4F_3\left(\frac{13}{30},\frac{19}{30},\frac{31}{30},\frac{37}{30};\frac{4}{3},\frac{4}{3},\frac{5}{3};\frac{25000 \xi ^3}{27}\right)  \\
			& = \xi ^2+\frac{283309 \xi ^5}{2592}+\frac{248945034845 \xi ^8}{6718464} +\mathcal{O}\left(\xi ^{11}\right) \\  \\
	 \Pi_{1'}(\xi)& = -\frac{1}{60 \sqrt{3} \ 5^{2/3} \pi } \ G_{4,4}^{2,4}\left(\frac{25000 \xi ^3}{27}\ \Bigg|
\begin{array}{c}
 \frac{13}{30},\frac{19}{30},\frac{31}{30},\frac{37}{30} \\
 \frac{1}{3},\frac{1}{3},0,\frac{2}{3} \\
\end{array}  
\right)  \\ &= \log \left( \frac{\xi}{2^{5/3}} \right) \ \Pi_1(\xi) + \left[ -\xi + \frac{10865 \xi ^4}{192}+\frac{78046343 \xi ^7}{4608}+ \mathcal{O} \left( \xi^{10} \right) \right]~, 
\label{fundsys}
\end{gleichung}\noindent
these transcendental numbers originate from the linear combinations for the $B$-period and not from the Picard-Fuchs equation itself. In particular, the $A$- and $B$-period are identified as follows
\begin{gleichung}
	\nu^{(0)}(\xi)	&=\Pi_1(\xi) \\
	 \nu_D^{(0)}(\xi)	&= \Pi_{1'}(\xi)+   \frac{2^{1/3}\sqrt{\pi }~ \Gamma \left(\frac{2}{3}\right)}{7~\Gamma \left(\frac{7}{6}\right)} \  \Pi_0(\xi) - \frac{7\pi~   \Gamma \left(-\frac{1}{3}\right)}{9~ \Gamma \left(-\frac{1}{6}\right) \Gamma \left(\frac{5}{6}\right)}  \ \Pi_2(\xi)~. \label{eq:class_per_quintic_lincomb}
\end{gleichung}\noindent
These are essentially the only transcendental numbers appearing in this context as the coefficients at higher orders in $\hbar$ come  from the classical expressions using \eqref{diffop_quintic}.

Picard-Fuchs equations for the $A$- and $B$-period at the first few orders in $\hbar$ are summarized in appendix \ref{sec_highpfops}.


\subsubsection{Quantum Free Energies from Quantum Mechanics}
As for the sextic oscillator we can compute quantum free energies. For the quintic we find
\begin{gleichung*}
	F_0(\nu) &= \left(	-\frac{3}{4} +  \frac{1}{2}\log \left(\frac{\nu}{2^{5/3}}	\right)	\right)\nu^2	+\frac{2^{1/3} \sqrt{\pi } \  \Gamma \left(\frac{2}{3}\right)}{7\ \Gamma \left(\frac{7}{6}\right)}\nu	-\frac{7 \pi  \ \Gamma \left(-\frac{1}{3}\right)}{27\ \Gamma \left(-\frac{1}{6}\right) \Gamma \left(\frac{5}{6}\right)}\nu^3	\\
	&\hspace{6cm}		+\frac{935 \sqrt{\pi }\ \Gamma \left(\frac{2}{3}\right)}{\num{1296}\cdot 2^{2/3}\ \Gamma \left(\frac{7}{6}\right)}\nu^4	+\frac{\num{2173}}{192} \nu ^5	+\mathcal O(\nu^6)\\
	F_1(\nu) &= -\frac{1}{24}\log \nu -\frac{11 \sqrt{\pi } \  \Gamma \left(\frac{1}{3}\right)}{72\cdot 2^{1/3}\ \Gamma \left(\frac{5}{6}\right)}\nu	+\frac{385 \sqrt{\pi } \ \Gamma \left(\frac{2}{3}\right)}{48\cdot 2^{2/3}\ \Gamma \left(\frac{1}{6}\right)}\nu^2	+\frac{\num{132245}}{\num{3456}} \nu ^3	\\
	&\hspace{6cm}	-\frac{\num{8527015} \pi  \ \Gamma \left(-\frac{1}{3}\right)}{\num{186624}\ \Gamma \left(-\frac{1}{6}\right) \Gamma \left(\frac{5}{6}\right)}\nu^4	+\mathcal O(\nu^5)\\
		F_2(\nu) &= -\frac{7}{\num{5760}}\frac{1}{ \nu ^2}	+\frac{\num{21171}}{\num{1024}}\nu	-\frac{\num{3882739} \sqrt{\pi } \ \Gamma \left(\frac{1}{3}\right)}{\num{165888} \cdot 2^{1/3}\ \Gamma \left(\frac{5}{6}\right)}\nu^2	\\
	&\hspace{2cm}	-\frac{\num{1599397248} \cdot 2^{1/3} \sqrt{\pi } \ \Gamma \left(-\frac{7}{3}\right)}{\num{1708143528275}\ \Gamma \left(-\frac{59}{6}\right)}\nu^3	+\frac{\num{3183085423}}{\num{73728}} \nu ^4		+\mathcal O(\nu^5) 
\end{gleichung*}\noindent
\begin{gleichung}
		F_3(\nu) &= \frac{31}{\num{161280}}\frac{1}{ \nu ^4}	+\frac{\num{3212061144326144} \sqrt[3]{2} \sqrt{\pi }\  \Gamma \left(-\frac{31}{3}\right)}{\num{49809465284499}\ \Gamma \left(-\frac{65}{6}\right)}\nu	+\frac{\num{275141423}}{\num{8192}}\nu^2\\
		&\hspace{2cm}	+\frac{\num{4068991173768}\ 2^{2/3} \sqrt{\pi }\ \Gamma \left(-\frac{11}{3}\right)}{\num{351738173908023175}\ \Gamma \left(-\frac{85}{6}\right)}\nu^3 \\
		&\hspace{2cm}		+\frac{\num{75795983236328485900} \sqrt[3]{2} \sqrt{\pi }\  \Gamma \left(-\frac{31}{3}\right)}{\num{98404065562059}\ \Gamma \left(-\frac{65}{6}\right)} \nu ^4		+\mathcal O(\nu^5)
\label{f1quintic}
\end{gleichung}

\noindent
which involve rational coefficients as well as transcendental ones. The structure of these transcendental numbers is inherited from the WKB periods. %
Topological string free energies are now highly constrained by this transcendental nature. It is one important step to recover these numbers in the string free energies. %
The leading singular coefficients of the free energies are the same as predicted from the gap condition \eqref{gapcondition}. 

\subsubsection{Ansatz for $F_1$}
\label{subsubsec_quinticF1}
For the topological string computation one necessary ingredient is a suitable ansatz for $F_1$. According to equation \eqref{ansatzfreeenergies} we make an ansatz in terms of the discriminant \eqref{eq:quintic_discriminant}. Using the classical mirror map we can compare this ansatz to $F_1$ in \eqref{f1quintic}. Unfortunately we obtain
\begin{gleichung}
	F_1^\text{ansatz}(\nu) &= -\frac{k_1}{24} (1+\alpha)\log \nu		\\
	 &-	k_1\left(	\frac{105 \alpha  }{128}+\frac{\num{408505} }{\num{10368}}	\right)\nu^3	-k_1\left(	\frac{\num{692255} \alpha}{\num{4096}}+\frac{\num{141101961685}}{\num{8957952}}	\right)\nu^6	+\mathcal O(\nu^9)~,
\label{ansatzwrong}
\end{gleichung}\noindent
which can not fit with the WKB result. In particular, there are no powers of $\nu$ which have transcendental coefficients as in \eqref{f1quintic}. Therefore, the ansatz \eqref{ansatzfreeenergies} together with the classical mirror map can not reproduce WKB computations.

An ad hoc generalization of the form
\begin{gleichung}
	F_1^\text{ansatz}(\nu)	=	p_1(\xi) + \alpha\log \left(p_2(\xi)\right)
\label{adhoc}
\end{gleichung}\noindent
where $p_1$ and $p_2$ are polynomials in $\xi$ and $\alpha$ is a constant does not seem to work out. 
We tested this ansatz up to polynomials of order four and found no solution. This mismatch in the transcendental coefficients 
might suggest that one has to consider a system with monodromy in ${\rm SL}(2,\mathbb{R})$ rather than the  monodromy in 
${\rm SL}(2,\mathbb{Z})$ that arose in the  problem with the symmetric sextic. Experience with the solution of the holomorphic anomaly for generic genus two hyperelliptic families~\cite{Klemm:2015iya} implies  that the problem 
can be solved after a further deformation and strongly suggests that the transcendental coefficients come from the 
restriction to the sub-slice. We consider such a deformation in the next section and show that all quantum periods 
can be characterized  by systems of two parameter differential  operators ${\cal D}_{2n}$. However, comparing this
result with a restriction of a solution of the genus two holomorphic anomaly equation is complicated and will
be deferred to future work.  Of course, such deformed quantum  mechanical problems are in itself very interesting as 
they can for example exhibit competing vacua that will lead to new non-perturbative effects.       

We have been informed by M. Mari\~no that at the Argyres-Douglas point\footnote{Such points have been studied in the application to quantum mechanics in \cite{Grassi:2018spf}.} of ${\rm SU}(5)$ $\mathcal N=2$ Yang-Mills theory one could obtain $F_1$ from the restriction of a genus four curve.


\subsection{A Two-parameter Family of Quintic Curves}
\label{subsec:perturbation}

In this section we enhance our discussion of quantum periods to a true  two-parameter higher-genus case\footnote{Recall that due to the rescaling symmetry \eqref{eq:rescaling} the parameter $\mathfrak g$ accompanying leading monomials in \eqref{eq:potential_class} did not represent a true modulus of the WKB curve.}. %
 The evaluation of hyperelliptic integrals becomes involved and laborious once more generic higher-genus curves are considered, which are \textit{not} a multi-cover of an elliptic curve. This gets severe once (1.) quantum corrections are to be computed or (2.) integrands depend on more parameters than just the energy. We show that nevertheless our proposed formalism applies with minimal modification, thus highlighting its true strength.
 From a physics point of view this makes it possible to investigate systems very different in their classical behavior, i.e., the number of potential wells and thus oscillatory trajectories, on the common footing of their WKB quantum periods.
 Last but not least, we expect the embedding of the one-parameter quintic family \eqref{defquintic} into a suitable two-parameter family to be a necessary step for the yet pending direct integration of the holomorphic anomaly recursion in case of a true genus-two geometry.

For concreteness, we study a parametric quintic potential with WKB curve
\begin{gleichung}
	\Sigma^{(5^\prime)}: \quad y^2	=	2\xi -x^2 +2x^5 + \eta (-x^2+2x^4)~,
\label{defpertquintic}
\end{gleichung}\noindent
where the coefficient of the leading monomial is again absorbed upon suitable rescaling. The perturbation is given by a quartic potential. The discriminant of $\Sigma_{5^\prime}$ turns out to be
\begin{gleichung}
	\Delta_{5^\prime}(\xi, \eta)		&=	\xi\left[	\num{25000}\xi^3+(1+\eta)^4(-27-27\eta+8\eta^3)+64\xi^2\eta^2(-125-125\eta+32\eta^3) \right. \\
			& \qquad \left.-4\xi\eta(1+\eta)^2(-225-225\eta+64\eta^3)	\right]~.
\end{gleichung}\noindent
Note that still an explicit factorization according to \eqref{eq:disc_factorization} can be given, as the criticial point condition is a quartic equation solvable in terms of radicals.

\begin{figure}[t]
\centering
        \includegraphics[width=0.55\textwidth]{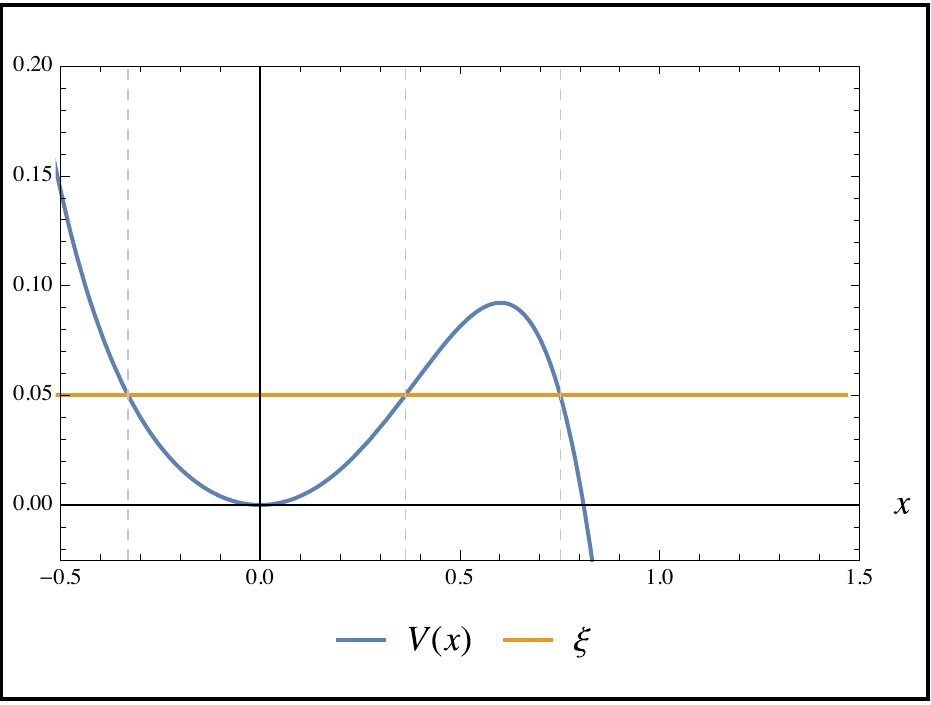}
        \caption{Perturbed quintic for $\eta=\frac{1}{10}$}
        \label{fig:pertquintic1}
\end{figure}

\begin{figure}[h]
\centering
        \includegraphics[width=0.55\textwidth]{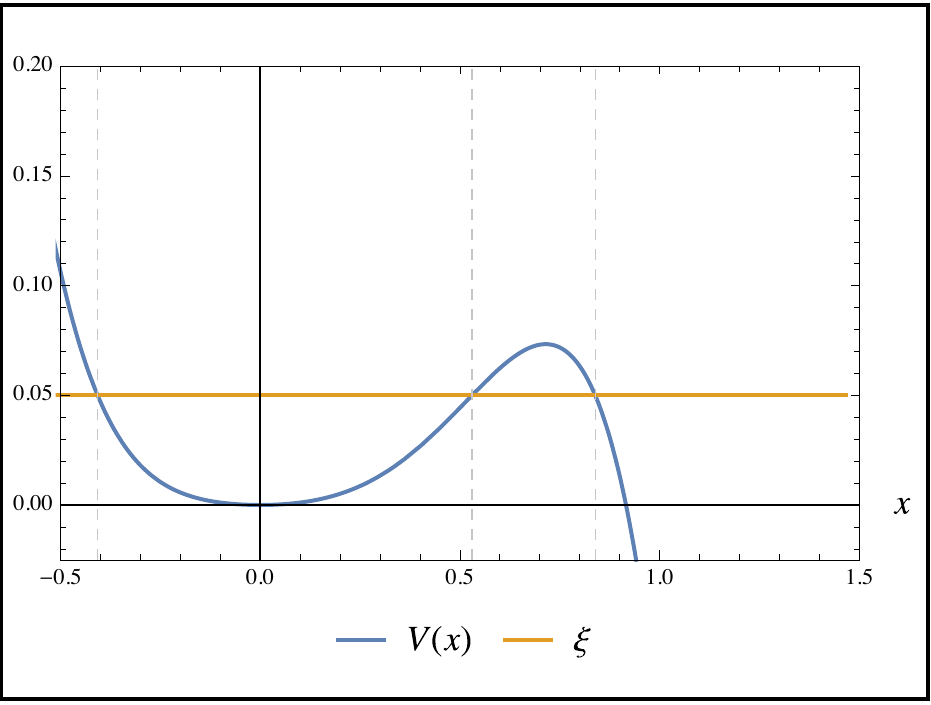}
        \caption{Perturbed quintic for $\eta\approx0.4$}
        \label{fig:pertquintic2}
\end{figure}

\begin{figure}[h]
\centering
        \includegraphics[width=0.55\textwidth]{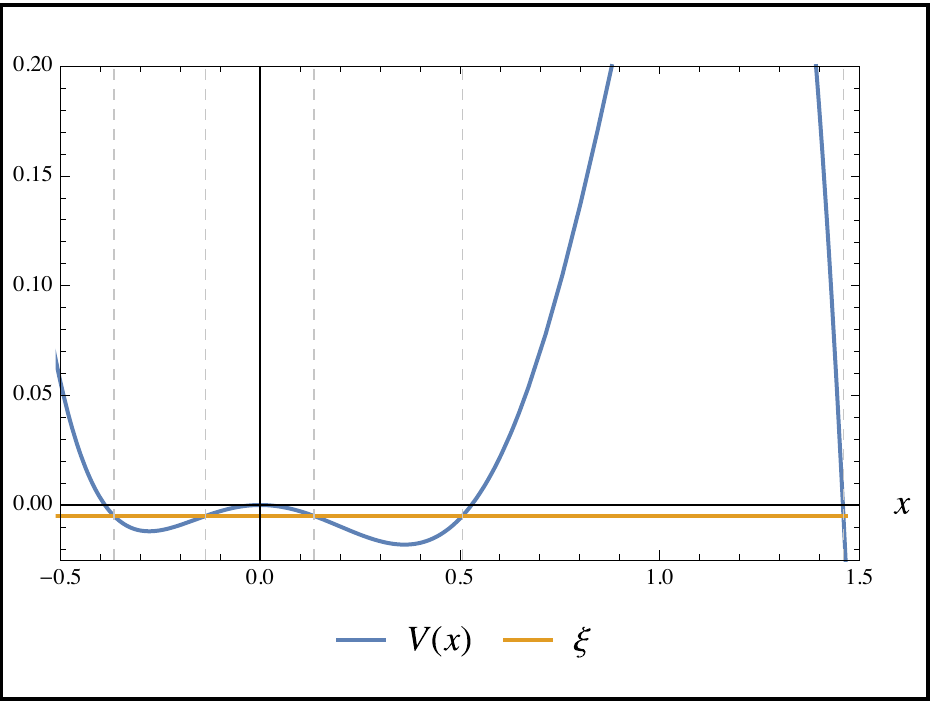}
        \caption{Perturbed quintic for $\eta=\frac{4}{5}$}
        \label{fig:pertquintic3}
\end{figure}

The perturbed quintic potential is visualized for different values of the perturbation parameter $\eta$ in Figs. \ref{fig:pertquintic1} - \ref{fig:pertquintic3}. Clearly, the number of real extrema changes from two to four as $\eta$ is varied. For $\eta$ sufficiently large one period integral previously belonging to a homology cycle around complex conjugated branch points now describes a physical action (at the classical level and for $\xi$ as in Fig. \ref{fig:pertquintic3}).

The Picard-Fuchs operator $\mathcal L$ generalizes in the two-parameter case to an ideal of Picard-Fuchs operators annihilating the periods. For the construction of operators generating the Picard-Fuchs ideal the ansatz in \eqref{eq:pfe_cohom_ansatz} gets extended including additional derivatives with respect to the second modulus $\eta$. Then the same procedure goes through, i.e., one subsequently eliminates monomials such that in the end one obtains the differential operator by imposing vanishing of the coefficients in front of the remaining monomials. In the two-parameter case there is an ambiguity in the vanishing condition. Independent choices of the free parameters yield a set of differential operators generating the Picard-Fuchs ideal. 

For the perturbed quintic curve \eqref{defpertquintic} there are two different Picard-Fuchs operators corresponding to two choices in the remaining parameters in the ansatz. Setting them separately to unity we find
\begin{gleichung*}
	\mathcal L_1	&=	\left[	-2 \xi  \left(320 \eta  \left(32 \eta ^3-75 \eta -50\right) \xi ^2+\left(56 \eta ^3+24
   \eta ^2-135 \eta -135\right) (\eta +1)^3 \right.\right. \\
   				&\quad\left.\left. +4 \left(-384 \eta ^5-640 \eta ^4+644 \eta
   ^3+2025 \eta ^2+1350 \eta +225\right) \xi \right)	\right]\partial_\xi^2	\\
				&	+\left[-2 \left(24 \eta ^7+\eta ^6 (96-512 \xi )+\eta ^5 \left(2048 \xi ^2-1024 \xi
   +63\right)+\eta ^4 (1288 \xi -309)\right.\right. \\
   				&\quad \left.\left.+\eta ^3 \left(-8000 \xi ^2+5400 \xi -786\right)-10
   \eta ^2 \left(800 \xi ^2-540 \xi +81\right)+45 \eta  (40 \xi -9)-81\right)\right]\partial_\eta\partial_\xi	\\
				&	+\left[-256 \eta  \left(4 \eta ^3-25 \eta -25\right) \xi ^2-3 \left(8 \eta ^3-27 \eta -27\right)
   (\eta +1)^3 \right. \\
   				&\quad \left.+8 \left(56 \eta ^5+64 \eta ^4 -202 \eta ^3-465 \eta ^2-300 \eta -45\right)
   \xi	\right]\partial_\xi \\
   				&	+\left[	8 \eta  \left(8 \eta  \left(16 \eta ^3-55 \eta -55\right) \xi -(\eta +1)^2 \left(8 \eta
   ^3-27 \eta -27\right)\right)	\right]\partial_\eta\\
   				&	+\left[	224 \eta  \left(-16 \eta ^3+25 \eta +10\right) \xi +14 \left(16 \eta ^3-24 \eta
   -9\right) (\eta +1)^2	\right]
\end{gleichung*}\noindent
\begin{gleichung}
	\mathcal L_2	&= \left[	-2 \xi  \left(160 \left(352 \eta ^4+336 \eta ^3-650 \eta ^2-775 \eta -75\right) \xi
   ^2+\left(288 \eta ^4+664 \eta ^3+12 \eta ^2 \right.\right.\right. \\
   				&\quad\left.\left.\left.-999 \eta -675\right) (\eta +1)^2-4
   \left(2032 \eta ^5+5256 \eta ^4+1118 \eta ^3-7721 \eta ^2-7350 \eta -1575\right) \xi
   \right)	\right]\partial_\xi^2 \\
   				&+\left[	-6 \left(128 \eta  \left(16 \eta ^4+8 \eta ^3-75 \eta ^2-125 \eta -50\right) \xi
   ^2+\left(32 \eta ^4+40 \eta ^3-108 \eta ^2 \right.\right.\right. \\
   				&\quad\left.\left.\left. -243 \eta -135\right) (\eta +1)^3-8 \left(80
   \eta ^6+184 \eta ^5-178 \eta ^4-962 \eta ^3-1115 \eta ^2-480 \eta -45\right) \xi
   \right)	\right]\partial_\eta\partial_\xi \\
   				&+\left[	8 \eta  (2 \eta +3) \left(8 \eta  \left(16 \eta ^3-55 \eta -55\right) \xi -(\eta +1)^2
   \left(8 \eta ^3-27 \eta -27\right)\right)	\right]\partial_\eta^2 \\
   				&+\left[	-64 \left(128 \eta ^4-96 \eta ^3-800 \eta ^2-875 \eta -75\right) \xi ^2-3 \left(32 \eta
   ^4+40 \eta ^3-108 \eta ^2 \right.\right. \\
   				&\quad\left.\left.-243 \eta -135\right) (\eta +1)^2+8 \left(256 \eta ^5+320
   \eta ^4-1052 \eta ^3-2613 \eta ^2-1905 \eta -360\right) \xi	\right] \xi\partial_\xi \\
   				&+\left[	-70 (2 \eta +3) \left(8 \left(16 \eta ^3-7 \eta -1\right) \xi -(\eta +1)^2 \left(8 \eta
   ^2-3\right)\right)	\right]~.
\end{gleichung}

With an obvious extension of the Frobenius method allowing for a double power series ansatz in \eqref{eq:pfe_ansatz} and \eqref{eq:pfe_ansatz_log} and possibly logarithms in the new parameter $\eta$ the periods are given by three pure power series 
\begin{gleichung}
	\Pi_0(\xi,\eta)	&=	1-\frac{7}{8}\eta^2 - \frac{245}{648}\eta^3 + \mathcal O\left(\eta^4\right) \\
				&\quad	+	\left(\frac{7}{9}\eta - \frac{35}{108}\eta^2+\frac{35}{162}\eta^3 +\mathcal O\left(\eta^4\right) \right)\xi	\\
				&\quad	+\left(-\frac{245}{72} + \frac{\num{3185}}{432}\eta -\frac{\num{54355}}{\num{5184}}\eta^2 +\mathcal O\left( \eta^3\right)\right)\xi^2	\\
				&\quad	+\left(\frac{\num{6545}}{648}-\frac{\num{271565}}{\num{3888}}\eta+\frac{\num{10417015}}{\num{46656}}\eta^2 + \mathcal O\left( \eta^3\right)\right)\xi^3+\mathcal O\left(\xi^4\right)	
\\
%
	\Pi_1(\xi,\eta)	&=	\left(1-\frac{1}{2}\eta+\frac{3}{8}\eta^2-\frac{5}{16}\eta^3+\mathcal O\left(\eta^4\right) \right)\xi	\\
				&\quad	+ \left(	\frac{3}{2}\eta-\frac{15}{4}\eta^2+\frac{105}{16}\eta^3+\mathcal O\left(\eta^4\right) \right)	\xi^2 \\
				&\quad	+\left(	\frac{35}{4}\eta^2-\frac{315}{8}\eta^3	+\mathcal O(\eta^4)\right)\xi^3 \\
				&\quad	+\left(	\frac{315}{16}-\frac{\num{3465}}{32}\eta+\frac{\num{45045}}{128}\eta^2+\mathcal O(\eta^3)\right)\xi^4	+\mathcal O\left(\xi^5\right)
\\
%
	\Pi_2(\xi,\eta)	&=	\eta + \frac{5}{6}\eta^2-\frac{5}{72}\eta^3 + \mathcal O\left(\eta^4\right) \\
				&\quad	+\left(	-\frac{5}{18}\eta^2+ \frac{35}{108}\eta^3+ \mathcal O\left(\eta^4\right)\right)\xi	\\
				&\quad	+\left(	\frac{35}{12}-\frac{455}{72}\eta+\frac{\num{8645}}{864}\eta^2 + \mathcal O\left( \eta^3\right)\right)\xi^2	\\
				&\quad	+\left(	\frac{\num{8645}}{324}\eta-\frac{\num{216125}}{\num{1944}}\eta^2+\frac{\num{6699875}}{\num{23328}}\eta^3 + \mathcal O\left( \eta^4\right)\right)\xi^3	\\
				&\quad	+\left(	\frac{\num{6699875}}{\num{23328}}\eta^2-\frac{\num{247895375}}{\num{139968}}\eta^3 + \mathcal O\left( \eta^4\right)\right)\xi^4		+\mathcal O\left(\xi^5\right)	\\
\end{gleichung}
and a logarithmic solution
\begin{gleichung}
	\Pi_{1^\prime}(\xi,\eta)	&=	\Pi_1(\xi,\eta)\cdot\log\left(	\frac{\xi}{2^{5/3}}	\right)	+\left(	\frac{1}{3}\eta^2+\frac{1}{6}\eta^3+\mathcal O\left(\eta^4\right)	\right)	\\
				&\quad	+\left(	-1-\frac{7}{6}\eta+\frac{31}{24}\eta^2+\mathcal O\left(\eta^3\right)	\right)\xi	\\
				&\quad	+\left(	\frac{53}{12}\eta-\frac{325}{24}\eta^2+\frac{\num{2575}}{96}\eta^3+\mathcal O\left(\eta^4\right)	\right)\xi^2		+ \mathcal O\left(\xi^3\right)~.
\end{gleichung}\noindent
As a consistency check setting the perturbation parameter $\eta$ to zero restores the old fundamental system \eqref{fundsys}.

As in the one-parameter case it is possible to construct quantum operators which applied to the classical periods give the quantum periods. In the two-parameter case these quantum differential operators have some freedom. In the construction \eqref{eq:quantum_diff_op} we used that the cohomology group $H^1(\Sigma,\mathbb C)$ is generated by derivatives of the differential $y(\xi)~\mathrm dx$ with respect to the modulus $\xi$. For two-parameter curves the cohomology group $H^1(\Sigma,\mathbb C)$ still has the same dimension $2g$ and can be generated by $y(\xi,\eta)~\mathrm dx$ and derivatives with respect to $\xi$, $\eta$ or combinations of both derivatives. This gives a freedom in writing down quantum differential operators $\hbar^{2n}\mathcal{D}_{2n}$ for multi-moduli curves. We prefer choosing derivatives with respect to the modulus $\xi$ only, since as a second consistency check we can then take the limit $\eta \rightarrow 0$ giving back the old quantum differential operators \eqref{diffop_quintic}. The first operator $\hbar^2\mathcal D_2$ is exemplarily written down in appendix \ref{sec_diffops}.

In this new setting of two-parameter curves it would be interesting to analyse how one could recover the WKB periods, in particular, the transcendental numbers or a closed expression for $F_1$. A guess could be that transcendental numbers arise from summing up contributions in the new parameter if one restricts to special limits in $\eta$.
 Furthermore, tuning $\eta$ could transform different quantum mechanical models into each other. For example taking $\eta=\frac{4}{5}$ transforms the quintic anharmonic oscillator potential to a potential of the form $\frac{9}{10}x^2-\frac{4}{5}x^4-x^5$ having four extrema, see Fig. \ref{fig:pertquintic3}. In the language of the curve this would yield five branch points located on the real axis. Constructing a symplectic basis of cycles on a genus two surface is well understood, for example as shown in Fig. \ref{fig:branch_cuts}. In this construction it is not necessary that the branch points lie on the real axis. For the quantum mechanical interpretation it makes a significant difference because these cycles encircle the classically allowed or forbidden regions. Additional allowed or forbidden regions can be interpreted as additional vacua and eventually tunneling contributions have to be taken into account. Perhaps in the WKB framework one has to regard more periods which are not considered so far. With our methods it is feasible to compute such periods but future work is required to give them a proper quantum mechanical interpretation. In this perspective analyzing the transition between anharmonic oscillators and more general potentials of different shapes could perhaps extend the theory of one-dimensional quantum mechanical systems, for instance, to generalized quantization conditions as proposed in \cite{ZinnJustin:2004ib,ZinnJustin:2004cg,Jentschura:2010zza,ddp97}. 





\section*{Acknowledgements}
We would like to thank Hans Jockers and Thorsten Schimannek  for useful conversations. Special thanks to Marcos Mari\~no for sharing his insights into the subject during collaboration in the initial state of the project.  
We further thank the Bonn-Cologne Graduate School of Physics and Astronomy (BCGS) for financial support.  F.F. also thanks the Studienstiftung des deutschen Volkes for support and A. K. thanks the  MSRI in Berkeley for hospitality during the final stage of this work.

\appendix

\section{Residues of Abelian Differentials on the WKB Curve}
\label{sec:app_res}

This appendix provides auxiliary residue calculations on the WKB curve $\Sigma \ : \ y^2 = 2\xi - x^2 + 2 x^d, d \geq 3$ (taken to be non-degenerate).
%
%
All WKB differentials are linear combinations of one-forms
\begin{equation}
\omega^m_k = \frac{x^m}{y^k} \dd x, \qquad m,k\in\mathbb{Z},
\end{equation}
so tentative residues are located at infinity or branch points.
We begin with points at infinity and discuss thes case $d=2g+2$ even and $d=2g+1$ odd separately.%
\paragraph{Case $d$ even.} If $d=2g+2$, there are two points at infinity $\infty^\pm \in \Sigma$, distinguished by the condition $y/x^{g+1}\rightarrow \pm 1$ as $x \rightarrow \infty$. At those points a local parameter is given by $z=1/x$, so
\begin{align}
\omega^m_k  = (-1)\frac{(\pm 1)^k}{z^{m+2-kd/2}} \left( 2 - z^{d-2}+ 2\xi \ z^{d}\right)^{-k/2} \ \dd z .
\end{align}
The sign $\pm 1$ refers to the point $\infty^\pm$ respectively. 
%
To compute the residue one uses the absolutely convergent binomial series (where $|\chi|<1$ and $ \alpha \in \mathbb{C}$)
\begin{align}
(1+\chi)^{\alpha} = \sum_{s=0}^\infty \binom{\alpha}{s} \ \chi^s, \qquad \binom \alpha s =
\begin{cases}\frac{\alpha (\alpha - 1)(\alpha - 2) \dotsm (\alpha - (s - 1))}{s!}, &\text{if }\quad s>0\\
1, &\text{if } \quad s=0\\
0, &\text{if }\quad  s<0
\end{cases}
\end{align}
for the case $\chi= \xi z^{d} - 1/2 \ z^{d-2}$ and $\alpha = -k/2$. 
%
In summary, the necessary condition for non-vanishing residue is
\begin{equation}
 m+1-kd/2 \in \lbrace 0,2,4, ... \rbrace
\end{equation}
and in this case one finds
\begin{equation}\label{eq:omega_res_even}
 (\pm 1)^k \ \mathrm{Res}_{\infty^\pm} \ \omega^m_k = 
 - 2^{-\frac{k}{2}} \sum_{s=0}^{\lfloor s_+ \rfloor} \  \binom{-k/2}{s} \ \binom{s}{l_s} \ (-2)^{l_s - s}  \  \xi^{l_s} 
\end{equation}
with
\begin{equation}
s_+ := \frac{m+1-kd/2}{d-2} \quad \text{and} \quad l_s := \frac{d-2}{2} (s_+ - s).
\end{equation}
The necessary condition is only satisfied for $n=0$, as $k=3n-1$ and $m\leq n(d-1)$ for the WKB differentials.

\paragraph{Case $d$ odd.} For odd $d=2g+1$ there is only one point at infinity $\infty \in \Sigma$ together with a local parameter $z=1/\sqrt{x}$.
%
%
The residue at infinity is given by
\begin{equation}\label{eq:omega_res_odd}
\ \mathrm{Res}_{\infty} \ \omega^m_k = 
 - 2^{1-\frac{k}{2}} \sum_{s=0}^{\lfloor s_+ \rfloor}\ \binom{-k/2}{s} \ \binom{s}{l_s} \ (-2)^{l_s - s} \ \xi^{l_s} \ \mathds{1}_{\mathbb{N}}(l_s)
\end{equation}
with
\begin{equation}
\mathds{1}_{\mathbb{N}}(l) := 
 \begin{cases}
    1,       & \quad \text{if } l \in \lbrace 0,1,2,... \rbrace\\
    0,  & \quad \text{otherwise}.\\
  \end{cases}
\end{equation}
%
Necessary conditions for a non-zero residue are
\begin{equation}
k \text{ even}\qquad \text{and} \qquad m+1-kd/2 \geq 0.
\end{equation}
For even $n$ the number $k=3n-1$ is odd, so $\mathcal{Q}_n$ has zero residue at infinity.

\paragraph{Behavior at branch points.} Turning to the behavior of $\omega^m_k$ at the $d$ finite branch points where $p^2(x)=0$, one has as local parameter $z=\sqrt{x - x_j}$. Thus
\begin{align}
\omega^m_k &= \frac{\left(z^2 + x_j\right)^m}{{\sqrt{\prod^{}_{\substack{i\neq j}}\left( z^2 + (x_j - x_i) \right) }}^{\ k} } \ \frac{2 }{z^{k-1}}\ \dd z
\end{align}
and the first factor is a holomorphic even function of $z$. Consequently if $k$ is odd, $\omega^m_k$ has zero residue at the branch points. Indeed, $k=3n-1$ is odd for terms in the WKB differentials $\mathcal{Q}_{n}$ with even $n$.

\section{Evaluation of Period Integrals}
\label{sec_evperiodint}

We have seen that the Picard-Fuchs equation easily yields expansions for period integrals in terms of their respective parameters up to arbitrarily high order. As the equation can be obtained from the differential and solved with modest effort, the only laborious point is to find the leading terms of the expansions in order to fix the correct linear combination of solutions.\footnote{As mentioned in section \ref{sec_geometry}, monodromy considerations already allow for partial identification of the periods.} Here the classical WKB periods $\oint_{A,B} y \ \dd x$ are defined by explicit specification of the homology cycle and shall be evaluated in the following.
%
Our approach applies to all hyperelliptic curves of the form
\begin{gleichung}
	y^2	=	2\xi - x^2 +2x^d \quad\text{with}\quad d>2~.
\label{gendefcurve}
\end{gleichung}\noindent
For $d>4$ this is particularly interesting as generically no closed expressions (say in terms of special functions) are known for the hyperelliptic integrals over $y(x,\xi)\ \dd x$.
%
For the sake of clarity we will focus on the case $d=5$.

Recall that the $A$- and $B$-cycle are defined such that they encircle pairs of branch points as given in Fig. \ref{fig:quintic_potential_a}.\footnote{In our convention branch cuts lie between branch points where $(\xi - V(x))$ is negative.}
There are no radical expressions for the branch points, which trace out trajectories in the complex plane as $\xi$ is being varied in an interval $I$.
As the value of the period does not change when replacing the integration contour by a larger one of the same homology class, we can choose a contour sufficiently large to enclose the full trajectories of the respective branch points, hence giving (locally) a $\xi$-independent integration contour.
 %
 %
  %
  %
 
  Consider $I=\left[0,v\right]$ with the root $v$ of $\Delta$. The series we will obtain for the $A$-period will be convergent in a disc $D_v(0)$, whereas the $B$-period will be convergent and single-valued in $D_v(0)\backslash \left[-v,0\right)$ due to the branch cut of the logarithm (which we take to lie on $\mathbb{R}_{<0}$).  %
  %
   For the quintic we find\footnote{Here we set $x_0= \frac{1}{3} \left(\sqrt[3]{2}-\sqrt[3]{\frac{5}{2}}-\frac{2}{\sqrt[3]{5}}\right)<0$.}
\begin{gleichung}
	\nu\hoch 0 (\xi)	&=	\frac{1}{\pi}\int_{x_0}^{5^{-1/3}} \sqrt{2\xi-x^2+2x^5}~\mathrm dx\\
	\nu_D\hoch 0 (\xi)	&=	-2i\int_0^{2^{-1/3}} \sqrt{2\xi-x^2+2x^5}~\mathrm dx~.
\label{expldef}
\end{gleichung}\noindent
The computation is based on the $\xi$-expansion of the integrand 
\begin{gleichung}
	y(x,\xi)	=	\sqrt{2\xi-x^2+2x^5}	= \sqrt{2x^5-x^2}	+	\frac{\xi}{\sqrt{2x^5-x^2}}	- \frac{\xi^2}{\sqrt{2x^5-x^2}} + \mathcal O(\xi^3)
\label{naive}
\end{gleichung}\noindent
which however is invalid for roots of $2x^5-x^2 = 0$ (an appropriate branch choice understood). As such roots lie within the integration range ($A$-period) or at its boundary ($B$-period), the range has to be split and a Cauchy principal value prescription to be used. 
In case of the $A$-period we obtain
\begin{gleichung}
	\nu\hoch 0(\xi)	=	\xi + \frac{315}{16}\xi^4 + \mathcal O(\xi^7)~.
\label{nulow}
\end{gleichung}\noindent
%
%
For the second intgeral in \eqref{expldef} we split the integration range into three parts
\begin{gleichung}
	\left[ 0,2^{-1/3} \right]	&=	\left[0,\epsilon\phantom{^{1}} \right]	+	\left[\epsilon, 2^{-1/3} - \delta\right] + \left[2^{-1/3}-\delta ,2^{-1/3}\right]
\label{splitting}
\end{gleichung}\noindent
where the middle part gives with the help of \eqref{naive}
\begin{gleichung}
	-2i\int_\epsilon^{2^{-1/3}-\delta} \sqrt{2\xi-x^2+2x^5}~\mathrm dx 	&=	\frac{\sqrt{\pi }~ \Gamma \left(\frac{5}{3}\right)}{2\cdot 2^{2/3} \Gamma \left(\frac{13}{6}\right)} +2\xi \log (\epsilon )-\frac{2}{3} \xi \log (2) \\
		& \quad -\frac{2\sqrt{\frac{2}{3}} }{3 \sqrt{\delta }}\xi^2-\frac{1}{2 \epsilon ^2}\xi^2-\frac{7 \pi~ \Gamma \left(-\frac{1}{3}\right)}{9~ \Gamma \left(-\frac{1}{6}\right)
   \Gamma \left(\frac{5}{6}\right)}\xi^2 + \mathcal O(\xi^3)~.
\label{middlepart}
\end{gleichung}\noindent
Here we have neglected higher order terms in $\epsilon$ and $\delta$ as they vanish for $\epsilon, \delta \rightarrow 0$ eventually. 
For the first integration part we do not use \eqref{naive} as it stands. Here, we first factorize $y^2=\prod_{i=1}^5(x-e_i(\xi))$ where the roots $e_i$ can be computed perturbatively in $\xi$. Since the integration variable $x$ is arbitrarily small for this integration range we expand the square root of three linear factors in $x$ and then perform the integration termwise. These three linear factors are picked by the condition that $e_i \nrightarrow 0$ as $\xi \rightarrow 0$. Expanding the intermediate result again in $\xi$ we obtain for the first part
\begin{gleichung}
	-2i\int_0^\epsilon \sqrt{2\xi-x^2+2x^5}~\mathrm dx 	&=-  (-\log (\xi )+2 \log (\epsilon )+1+\log (2))\xi +\frac{\xi ^2}{2 \epsilon ^2} + \mathcal O(\xi^3)~.
\label{firstpart}
\end{gleichung}\noindent
The same strategy can be applied to the upper integration 
domain. As it turns out, this part does not give any finite contributions to the final result. It merely cancels singular terms in  $\delta$.

Adding all three contributions we finally obtain the leading behaviour of the $B$-period
\begin{gleichung}
	-2i\int_0^\epsilon \sqrt{2\xi-x^2+2x^5}~\mathrm dx 	& \\ & \hspace*{-3.5cm} = \frac{\sqrt{\pi }\ \Gamma \left(\frac{5}{3}\right)}{2\cdot 2^{2/3}\ \Gamma \left(\frac{13}{6}\right)} - \left( 1 - \log \left(\frac{\xi}{2^{5/3}} \right)	\right)\xi 	-\frac{7 \pi\  \Gamma \left(-\frac{1}{3}\right)}{9\ \Gamma \left(-\frac{1}{6}\right) \Gamma \left(\frac{5}{6}\right)}\xi^2 + \mathcal O(\xi^3)~.
\label{bfull}
\end{gleichung}\noindent

For the higher order WKB periods one could do a similar computation. Fortunately, this is not necessary because the operators $\mathcal D_{2n}$ allow us to compute the quantum corrections directly from the classical periods.

\section{Picard-Fuchs Operators for WKB Periods}
\label{sec_highpfops}

In this appendix we collect the Picard-Fuchs operators annihilating the WKB periods for the quintic and sextic anharmonic oscillator. Results include the leading and the first three subleading orders corresponding to WKB periods of order $\lbrace \hbar^{0},\hbar^{2},\hbar^{4}, \hbar^{6}  \rbrace$. In the present notation the differential operator $\mathcal L_\text{PF}^{(2n)}$ annihilates the WKB periods $\nu^{(2n)}\hbar^{2n}$ and $\nu_D^{(2n)}\hbar^{2n}$.

On first sight it seems that in higher order Picard-Fuchs operators new singularities arise in terms of additional zeroes of the polynomial multiplying the highest derivative. However, these are spurious poles since writing the Picard-Fuchs equation in a coordinate centered at such a point one obtains a fundamental system spanned by regular solutions. This is in accordance with our geometric expectation: the radius of convergence of a solution is determined by the distance to the nearest degeneration point, which must be a root of the discriminant. Clearly, the respective discriminant appears in all Picard-Fuchs operators of the sextic and quintic periods.

\subsection*{Picard-Fuchs Operators for the Sextic Oscillator}
\vspace{0.25cm}

\begin{scriptsize}
\begin{gleichung}
	 \mathcal L_\text{PF}^{(0)}	&=	\left( {54} \xi^{2} {-} {1} \right)\  \xi^{}\partial_{\xi}^{4} \\
						& \quad + 2 \left( {162} \xi^{2} {-} {1} \right) \  \partial_{\xi}^{3} \\		
						& \quad + {354} \  \xi^{}\partial_{\xi}^{2} \\
						& \quad + {30} \  \partial_{\xi}^{} \\
						& \phantom{\hspace{12cm}}
\end{gleichung}\noindent
\begin{gleichung}
	\mathcal L_\text{PF}^{(2)}	&=	\left( {54} \xi^{2} {-} {1} \right) \left( {1782} \xi^{2} {-} {65} \right) \ \xi^{1}\partial_{\xi}^{3} \\
						& \quad + 3 \left( {224532} \xi^{4} {-} {11124} \xi^2 + 65 \right) \ \partial_{\xi}^{2} \\
						& \quad + 66 \left( {14418} \xi^{2} {-} {947} \right) \ \xi^{}\partial_{\xi}^{} \\
						& \quad + 96 \left( {1782} \xi^{2} {-} {185} \right) \\
						& \phantom{\hspace{12cm}}
\end{gleichung}\noindent
\begin{gleichung}
	\mathcal L_\text{PF}^{(4)}	&=	\xi  \left(54 \xi ^2-1\right) \left(7409920500 \xi ^4-330245820 \xi ^2+9632693\right)~\partial_\xi^3 \\
						& \quad +5 \left(880298555400 \xi ^6-47848497228 \xi ^4+1758643758 \xi ^2-9632693\right)~\partial_\xi^2 \\
						& \quad +30 \xi  \left(398653722900 \xi ^4-24604325820 \xi ^2+1016813357\right)~\partial_\xi \\
						& \quad +960 \left(7409920500 \xi ^4-486072540 \xi ^2+40410853\right) \\
						& \phantom{\hspace{12cm}}
\end{gleichung}\noindent
\begin{gleichung}
	\mathcal L_\text{PF}^{(6)}	&=	\xi  \left(54 \xi ^2-1\right) \left(1593291078484200 \xi ^6+86962831950060 \xi ^4+7919960534814 \xi
   ^2-50291624911\right)~\partial_\xi^3 \\
						& \quad +\left(1290565773572202000 \xi ^8+78238588651670880 \xi ^6+7864991012868984 \xi ^4\right. \\
						& \quad \left. \qquad -96630505323144 \xi^2+352041374377\right)~\partial_\xi^2 \\
						& \quad +\left( \xi  868343637773889000 \xi ^6+63008309812806540 \xi ^4+7170473076071550 \xi^2\right. \\
   						& \quad \left. \qquad -50170087365031\right)~\partial_\xi \\
						& \quad +30240 \left(177032342053800 \xi ^6+16620344819244 \xi ^4+1987953607118 \xi^2-28451410671\right) \\
						& \phantom{\hspace{12cm}}
\end{gleichung}\noindent
\end{scriptsize}

\subsection*{Picard-Fuchs Operators for the Quintic Oscillator}
\vspace{0.25cm}

\begin{scriptsize}
\begin{gleichung}
	 \mathcal L_\text{PF}^{(0)}	&=	 \left( {25000} \xi^{3} {-} {27} \right) \ \xi^{}\partial_{\xi}^{4} \\
						& \quad + \left( {200000} \xi^{3} {-} {54} \right) \ \partial_{\xi}^{3} \\		
						& \quad + {317500}   \ \xi^{2}\partial_{\xi}^{2} \\
						& \quad + {35000} \  \xi\partial_{\xi}^{} \\
						& \quad + \frac{6545}{2} \\
						& \phantom{\hspace{12cm}}
\end{gleichung}\noindent
\begin{gleichung}
	 \mathcal L_\text{PF}^{(2)}	&=	 2 \left( {32000000} \xi^{3} {+} {193347} \right) \left( {25000} \xi^{3} {-} {27} \right) \  \xi^{}\partial_{\xi}^{4}  \\
						& \quad + 8 \left( {2600000000000} \xi^{6} {+} {19118700000} \xi^{3}- 5220369 \right)  \ \partial_{\xi}^{3} \\		
						& \quad + 125000 \left( {573440000} \xi^{3} {+} {5525577} \right)  \ \xi^2 \partial_{\xi}^{2} \\
						& \quad + 18000 \left( {3656000000} \xi^{3} {+} {49908357} \right) \  \xi^{}\partial_{\xi}^{} \\
						& \quad + 267 995 \left( {32000000} \xi^{3} {+} {720657} \right)  \\
						& \phantom{\hspace{12cm}}
\end{gleichung}\noindent
\begin{gleichung}
	 \mathcal L_\text{PF}^{(4)}	&=	 \ 2\xi\left(	 25000 \xi ^3-27	\right)\left(	 33013760000000000 \xi ^6+1565984563200000 \xi ^3+6305572607481	\right)\ \partial_\xi^4  \\
						& \quad + 36\left( 825344000000000000000 \xi ^9+45674549760000000000 \xi ^6+203138823048300000 \xi ^3\right. \\
						&\qquad \left. -56750153467329	\right)\ \partial_\xi^3 \\
						& \quad +5000\left(	31726223360000000000 \xi ^6+2051693233459200000 \xi ^3+11526405450672231	\right) \  \xi ^2\ \partial_\xi^2 \\
						& \quad + 14000\left(	19890790400000000000 \xi ^6+1504793310105600000 \xi ^3+11080573290031581\right) \  \xi \ \partial_\xi \\
						& \quad + 1216215\left(	99041280000000000 \xi ^6+8399258009600000 \xi ^3+109484562862143	\right)  \\
						& \phantom{\hspace{12cm}}
\end{gleichung}\noindent
\begin{gleichung}
	 \mathcal L_\text{PF}^{(6)}	&=	 \ 2\xi\left(	 25000 \xi ^3-27	\right)\left(	 108190984437760000000000 \xi ^6-2201331444040108800000 \xi ^3 \right.  \\
	 					&\qquad\left.-1084714198114610277	\right)\ \partial_\xi^4 \\
						& \quad + 16\left(8790517485568000000000000000 \xi ^9-200225951261089740000000000 \xi ^6\right. \\
						&\qquad \left. -71323951693284191700000 \xi ^3+29287283349094477479	\right)\ \partial_\xi^3 \\
						& +35000\left(	33400102481428480000000000 \xi ^6-849330559112858956800000 \xi ^3 \right. \\
						&\qquad\left. -561279483265056745473	\right) \xi ^2\ \partial_\xi^2 \\
						& \quad + 14000\left(	255446642185011200000000000 \xi ^6-7321188595679726472000000 \xi ^3\right. \\
						&\qquad\left.-5333494905100446357357\right) \xi\ \partial_\xi \\
						& \quad + 210900515\left(	15455854919680000000000 \xi ^6-502837975009094400000 \xi ^3-627242552904521151	\right)  \\
						& \phantom{\hspace{12cm}}
\end{gleichung}\noindent
\end{scriptsize}


\section{Differential Operators for Quantum Corrections}
\label{sec_diffops}

As for the one-parameter models we compute differential operators which applied on the classical periods give the quantum corrections at a certain order in $\hbar$. In the two-parameter models these operators get quite messy. An example is given for $\hbar^2\mathcal D_2$

\begin{scriptsize}
\begin{gleichung}
\mathcal D_2	&=	\left[	\left(8 \eta ^3-25 \eta -25\right) \xi  \left(800000 \eta  \left(4 \eta ^3-15 \eta -15\right) \xi ^4-\left(-8 \eta ^3+27
   \eta +27\right)^2 (\eta +1)^6-64 \eta ^2 \left(768 \eta ^6-5584 \eta ^4 \right.\right.\right. \\
   			&\quad \left.\left.\left. -5584 \eta ^3+10125 \eta ^2+20250 \eta
   +10125\right) (\eta +1)^2 \xi ^2+12 \eta  \left(256 \eta ^6-1784 \eta ^4-1784 \eta ^3 \right.\right.\right. \\
   			&\quad\left.\left.\left.+3105 \eta ^2+6210 \eta
   +3105\right) (\eta +1)^4 \xi +8 \left(32768 \eta ^9-250880 \eta ^7-250880 \eta ^6+455000 \eta ^5+910000 \eta ^4 \right.\right.\right. \\
   			&\quad\left.\left.\left.+539375 \eta ^3+253125 \eta ^2+253125 \eta +84375\right) \xi ^3\right)	\right]/ \\
			&\quad\left[	-3072 \eta ^{14}+3072 \eta ^{13} (32 \xi -5)-192 \eta ^{12} \left(4096 \xi ^2-1536 \xi +17\right)-384 \eta ^{11} \left(2048
   \xi ^2+1592 \xi -349\right) \right. \\
   			&\quad\left. +96 \eta ^{10} \left(74880 \xi ^2-36736 \xi +3293\right)+96 \eta ^9 \left(14336 \xi ^3+149760
   \xi ^2-28044 \xi -171\right)+3 \eta ^8 \left(-4797440 \xi ^2 \right.\right. \\
   			&\quad \left.\left.+3367040 \xi -399669\right)-24 \eta ^7 \left(384000 \xi
   ^3+2697600 \xi ^2-992680 \xi +84801\right)-12 \eta ^6 \left(768000 \xi ^3 \right.\right. \\
   			&\quad \left.\left. +3631200 \xi ^2-926880 \xi +53467\right)+24
   \eta ^5 \left(640000 \xi ^3+2628800 \xi ^2-1129080 \xi +108297\right) \right. \\
   			&\quad\left. +6 \eta ^4 \left(5120000 \xi ^3+21168000 \xi
   ^2-8614464 \xi +799281\right)+24 \eta ^3 \left(640000 \xi ^3+3528000 \xi ^2-1701000 \xi \right.\right. \\
   			&\quad \left.\left. +175257\right)+3780 \eta ^2
   \left(5600 \xi ^2-4320 \xi +567\right)-68040 \eta  (40 \xi -9)+76545	\right]\cdot\partial_\xi^3 \\
	&\\
			&+\left[	-3 \left(-8 \eta ^3+27 \eta +27\right)^2 \left(16 \eta ^3-55 \eta -55\right) (\eta +1)^6+128000 \eta  \left(3712 \eta
   ^6-25400 \eta ^4-25400 \eta ^3 \right.\right. \\
   			&\quad\left.\left. +43125 \eta ^2+86250 \eta +43125\right) \xi ^4-32 \eta ^2 \left(106496 \eta ^9-1155840
   \eta ^7-1155840 \eta ^6+4149160 \eta ^5+8298320 \eta ^4 \right.\right. \\
   			&\quad\left.\left. -781715 \eta ^3-14792625 \eta ^2-14792625 \eta -4930875\right)
   (\eta +1)^2 \xi ^2+8 \eta  \left(22528 \eta ^9-235520 \eta ^7-235520 \eta ^6 \right.\right. \\
   			&\quad\left.\left.+818472 \eta ^5+1636944 \eta ^4-127203 \eta
   ^3-2837025 \eta ^2-2837025 \eta -945675\right) (\eta +1)^4 \xi +32 \left(655360 \eta ^{12} \right.\right. \\
   			&\quad\left.\left. -7360512 \eta ^{10}-7360512
   \eta ^9+25942400 \eta ^8+51884800 \eta ^7+1192400 \eta ^6-74250000 \eta ^5-88171875 \eta ^4 \right.\right. \\
   			&\quad\left.\left. -80437500 \eta ^3-83531250
   \eta ^2-55687500 \eta -13921875\right) \xi ^3	\right]/ \\
			&\quad\left[	-24576 \eta ^{14}+24576 \eta ^{13} (32 \xi -5)-1536 \eta ^{12} \left(4096 \xi ^2-1536 \xi +17\right)-3072 \eta ^{11}
   \left(2048 \xi ^2+1592 \xi -349\right) \right. \\
   			&\quad\left. +768 \eta ^{10} \left(74880 \xi ^2-36736 \xi +3293\right)+768 \eta ^9 \left(14336
   \xi ^3+149760 \xi ^2-28044 \xi -171\right)+24 \eta ^8 \left(-4797440 \xi ^2 \right.\right. \\
   			&\quad\left.\left. +3367040 \xi -399669\right)-192 \eta ^7
   \left(384000 \xi ^3+2697600 \xi ^2-992680 \xi +84801\right)-96 \eta ^6 \left(768000 \xi ^3 \right.\right. \\
   			&\quad\left.\left.+3631200 \xi ^2-926880 \xi
   +53467\right)+192 \eta ^5 \left(640000 \xi ^3+2628800 \xi ^2-1129080 \xi +108297\right) \right. \\
   			&\quad\left. +48 \eta ^4 \left(5120000 \xi
   ^3+21168000 \xi ^2-8614464 \xi +799281\right)+192 \eta ^3 \left(640000 \xi ^3+3528000 \xi ^2-1701000 \xi \right.\right. \\
   			&\quad\left.\left.+175257\right)+30240 \eta ^2 \left(5600 \xi ^2-4320 \xi +567\right)-544320 \eta  (40 \xi -9)+612360	\right]\cdot\partial_\xi^2 \\
	&\\
			&+\left[	5 \left(-3 \eta  \left(-8 \eta ^3+27 \eta +27\right)^2 (\eta +1)^5+640 \eta  \left(1568 \eta ^6-10700 \eta ^4-10700 \eta
   ^3 +18125 \eta ^2+36250 \eta  \right.\right.\right. \\
   			&\quad\left.\left.\left. +18125\right) \xi ^3+48 \eta ^2 \left(128 \eta ^6-872 \eta ^4-872 \eta ^3+1485 \eta ^2+2970
   \eta +1485\right) (\eta +1)^3 \xi   -12 \left(4096 \eta ^{10}+4096 \eta ^9\right.\right.\right. \\
   			&\quad\left.\left.\left. -19520 \eta ^8-39040 \eta ^7-27520 \eta ^6-24000
   \eta ^5+67875 \eta ^4+359500 \eta ^3+551250 \eta ^2+367500 \eta +91875\right) \xi ^2\right)	\right]/ \\
			&\left[	-2048 \eta ^{14}+2048 \eta ^{13} (32 \xi -5)-128 \eta ^{12} \left(4096 \xi ^2-1536 \xi +17\right)-256 \eta ^{11} \left(2048
   \xi ^2+1592 \xi -349\right) \right. \\
   			&\quad\left. +64 \eta ^{10} \left(74880 \xi ^2-36736 \xi +3293\right)+64 \eta ^9 \left(14336 \xi ^3+149760
   \xi ^2-28044 \xi -171\right)+2 \eta ^8 \left(-4797440 \xi ^2 \right.\right. \\
   			&\quad\left.\left. +3367040 \xi -399669\right)-16 \eta ^7 \left(384000 \xi
   ^3+2697600 \xi ^2-992680 \xi +84801\right)-8 \eta ^6 \left(768000 \xi ^3+3631200 \xi ^2 \right.\right. \\
   			&\quad\left.\left.-926880 \xi +53467\right)+16 \eta
   ^5 \left(640000 \xi ^3+2628800 \xi ^2-1129080 \xi +108297\right)+4 \eta ^4 \left(5120000 \xi ^3 \right.\right. \\
   			&\quad\left.\left. +21168000 \xi ^2-8614464
   \xi +799281\right)+16 \eta ^3 \left(640000 \xi ^3+3528000 \xi ^2-1701000 \xi +175257\right) \right. \\
   			&\quad\left. +2520 \eta ^2 \left(5600 \xi
   ^2-4320 \xi +567\right)-45360 \eta  (40 \xi -9)+51030	\right]\cdot\partial_\xi
\label{diffop_perturbed}
\end{gleichung}\noindent  
\begin{gleichung*}
\phantom{\mathcal D_2}		&+\left[	35 \left(32 \eta  \left(-8 \eta ^3+25 \eta +25\right)^2 \xi ^2+3 \eta ^2 \left(8 \eta ^3-27 \eta -27\right) (\eta +1)^4-3
   \left(512 \eta ^6-2960 \eta ^4-2960 \eta ^3 \right.\right.\right. \\
   			&\quad\left.\left.\left.+4125 \eta ^2+8250 \eta  +4125\right) (\eta +1)^2 \xi \right)	\right]/ \\
			&\quad\left[	-1024 \eta ^{14}+1024 \eta ^{13} (32 \xi -5)-64 \eta ^{12} \left(4096 \xi ^2-1536 \xi +17\right)-128 \eta ^{11} \left(2048
   \xi ^2+1592 \xi -349\right) \right. \\
   			&\quad\left. +32 \eta ^{10} \left(74880 \xi ^2-36736 \xi +3293\right)+32 \eta ^9 \left(14336 \xi ^3+149760
   \xi ^2-28044 \xi -171\right)+\eta ^8 \left(-4797440 \xi ^2 \right.\right. \\
			&\quad\left.\left. +3367040 \xi -399669\right)-8 \eta ^7 \left(384000 \xi
   ^3+2697600 \xi ^2 -992680 \xi +84801\right)-4 \eta ^6 \left(768000 \xi ^3+3631200 \xi ^2 \right.\right. \\
   			&\quad\left.\left.-926880 \xi +53467\right)+8 \eta
   ^5 \left(640000 \xi ^3+2628800 \xi ^2-1129080 \xi +108297\right)+2 \eta ^4 \left(5120000 \xi ^3+21168000 \xi ^2 \right.\right. \\
   			&\quad\left.\left. -8614464
   \xi +799281\right)+8 \eta ^3 \left(640000 \xi ^3+3528000 \xi ^2-1701000 \xi +175257\right)+1260 \eta ^2 \left(5600 \xi
   ^2-4320 \xi \right.\right. \\
   			&\quad\left.\left.  +567\right)-22680 \eta  (40 \xi -9)+25515	\right]~.
\end{gleichung*}\noindent
\end{scriptsize}

\newpage

\addcontentsline{toc}{section}{References}
\bibliographystyle{utphys}
\bibliography{pqm}
\end{document}